DRAFT September 24, 2001

# What Should be Hidden and Open in Computer Security: Lessons from Deception, the Art of War, Law, and Economic Theory

## Peter P. Swire[1]

Introduction

I.      Security and Obscurity

      A.      Security and Obscurity in the Physical World

      B.      The Weakness of Obscurity and Deception in Modern Computer Security

            1.      Firewalls

            2.      Packaged Software

            3.      Encryption

II.     A Model for What Should be Open and Hidden in Computer Security

      A.      First-Time and Repeated Attacks

      B.      Learning from Attacks: The Distinction Between Surveillance and Other Defenses

      C.      Communication Among Attackers and the Diffusion of Knowledge

      D.      A Dynamic Model and the Accountability Effects of Openness

            1.      The Security-Enhancing Effect

[1] Peter P. Swire is Professor of Law at the Ohio State University and Visiting Professor at the George Washington University Law School. From early 1999 until January, 2001 he served as the Clinton Administration's Chief Counsel for Privacy, in the Office of Management and Budget. In that position he worked extensively on encryption policy and government computer security issues, including the Federal Intrusion Detection Network (FIDNet), Carnivore, and critical infrastructure issues. In 2000 he chaired a 15-agency White House Working Group on how to update wiretap and electronic surveillance laws for the Internet age. The Administration's legislative proposal was introduced into the Senate in 2000 as S. 3083. Contact information: email pswire@law.gwu.edu; phone (301) 213-9587; web: www.osu.edu/units/law/swire.htm.

My thanks for informative comments on this project from people including Matt Blaze, Bill Bratton, David Brin, Robert Cottrell, Mark Lemley, Peter Neumann, and Steve Schooner. Remaining errors are my own. Thanks also for fine research assistance from John Kammerer, and research support from the George Washington University Law School.

[Disclaimer: This interim draft is submitted to the Telecommunications Policy Research Conference to meet the September 24 deadline. Comments emphatically welcome. I hope to post updated drafts to fill in sections that are not yet complete. Sections introduced by a bracket "[" are in especially preliminary form. Fuller citations will be provided in subsequent drafts, and the usefulness of the paper will increase if interested readers can suggest relevant literature to cite.]

Imagine a military base. It is defended against possible attack. Do we expect the base to reveal the location of booby traps and other defenses? No. The information about the defenses is hidden, closed, and definitely not put on the Internet.

Now imagine a web site on the Internet, or a company computer system that shares complex information with its customers and business partners. The system owner will have to disclose a good deal about the system -- about its hardware and software -- so that the owners of other systems know how to inter-operate with his system. Even more importantly, when he builds his system, he will want to know a lot about the software and hardware he is building into his own system -- he will consider his own system secure only if he trusts each component in the system.

This need to inter-operate with other people's software and hardware is one important reason why we would expect the optimal level of openness for computer security to be higher than for security of the military base. There are other important reasons, discussed below, for believing that the optimal level of openness is often higher for computer security than for traditional physical security. I believe this is an interesting and important question that has not been the focus of enough study to date. But it is an unfamiliar problem -- what should be open or hidden when it comes to computer security.

Fortunately, we can turn it into a more familiar problem. Think about the system owner who does not reveal the booby traps. That owner has monopoly information -- he or she knows more about the system's defenses than anyone else. By contrast, think about the system owner when the specs have been shared freely over the Internet with designers for other systems. In that case, we approach a competitive market, where the system owner does not necessarily have any advantage over outsiders when it comes to understanding system security. Often, outside experts will know more about the security vulnerabilities than Desi, who may be a small business owner without particular security expertise.

The unfamiliar problem of closedness or openness in computer security thus can be transformed into the familiar question of when to favor monopoly versus competition. In economic theory (in contrast to military strategy) there is a strong presumption on the side of competition and free flow of information. One exception to this presumption is in the realm of intellectual property, where there are major debates today over the conditions where a monopoly is appropriate (exclusive rights should be granted to a copyright or patent holder) or where a competitive approach is appropriate (fair use or similar rules should limit the exclusive control of the rights holder). One claim of this paper is that the familiar tools of economic analysis can illuminate the study of closedness and openness in computer security. An implication back for economic theory is to generalize the debate about the limits on free flow of information that economists have studied for intellectual property. There are important efficiency questions, in an era where computer security is becoming so important, about how to design the flow of information about the security of computer systems.

Let's return for a moment to the military base. We assumed initially that secretness and deception were important to military security. It turns out, however, that there are contrasting themes within the theory of military security. Notably, the ancient Chinese strategist Sun Tzu stated that "all warfare is based on deception." On the other hand, the famous German writer Clausewitz, in his comprehensive theory of military strategy, believed that deception played only a minor role. Clausewitz was far more focused on concentrating overwhelming force at the point of attack, what computer experts today call a "brute force attack." The contrast between Sun Tzu and Clausewitz shows that a longstanding debate has existed within the military realm about the usefulness of hiddenness versus the usefulness of focusing on fundamentally strong security (security that can withstand even knowledge by the enemy). Thus, there is a need for analysis of what should be hidden and open in security even in the military realm, where one might have expected hiddenness to hold sway.

Part I of the paper introduces the themes of deception and openness in physical world security, and shows how greater openness is typical for major computer security topics, such as firewalls, packaged software, and encryption. For all three of these categories, passwords are kept hidden but the underlying structure of the software is widely known outside of the realm of each owner who is defending his own computer system. Indeed, a slogan for many computer programmers is that "there is no security through obscurity," highlighting the idea that fundamentally strong security, tested by outside hackers, is the best route to computer security.

Part II provides a theoretical analysis of why openness is more often optimal for computer security than for physical security. Whereas in the physical world deception often works against a first attack, such as when the attacker falls into a hidden pit, the use of deception is far less likely to be effective against multiple attacks, where the attackers learn the location of the hidden pits. For areas including firewalls, packaged software, and encryption, system designers have to plan for the likelihood of repeated attacks, where the attackers learn from their earlier efforts, and where successful attacks by one person can be communicated to a large peer community of other possible attackers. Under these conditions, deception and hidden traps are especially ineffective, and security planning must proceed on the assumption of a good deal of openness about the system design. This Part's analysis of computer security presents strategies for system designers who face these sorts of attacks, notably by re-creating conditions that resemble a first-time attack.

Part III introduces the economic analysis of what should be hidden and open in computer security. This Part highlights the likelihood that companies with monopoly power will try to hide more about the workings of their system than is socially optimal. A related conclusion is that organizations that do not face a market test, notably government agencies, have strong incentives to keep secret more than is socially optimal about computer security.

Part IV examines the approaches of Sun Tzu and Clausewitz to the role of hiddenness and

-4-

deception in military strategy. The analysis shows that the usefulness of hiddenness varies with factors such as terrain, changing technology, and the nature of the attack. These factors, in turn, mirror the factors in computer security and economic theory that establish when hiddenness is likely to be effective or ineffective as a defense strategy.

Part V applies the insights from computer security, economic theory, and military strategy to a number of legal and policy issues. First, and most generally, the open source movement has proclaimed the importance of open code to effective computer security. The analysis in this paper is designed to answer the question of *when* there is security through obscurity, that is, when being either open or closed is likely to lead to the optimal level of security. Second and relatedly, the anti-circumvention provisions of the Digital Millennium Copyright Act set strict limits on the legality of publicizing flaws in some computer security products. The analysis here suggests new reasons for believing that the computer security exception to the DMCA is substantially too narrow, and explains why the exception creates possible hazards to military and other computer security. Third, the paper explores whether a common strategy for protecting military secrets -- granting security clearances and classifying information -- is appropriate for civilian computer security, and concludes that it generally is not. Fourth, the combined results of the first three applications shows the way that technology (conditions for open source security), law (the DMCA), and institutions (the classification system) can substitute for each other in creating an optimal degree of openness or hiddenness in computer security. Fifth, the analysis of computer security here shows that surveillance activities are affected far less in the shift from physical to computer security than other security measures. For computer surveillance, such as for the Carnivore e-mail system, an open source approach is therefore likely not justified on the basis of improved security alone, although an open approach may be justifiable for other reasons. Sixth, the entire analysis of what should be open or closed has important implications for the U.S. critical infrastructure protection program. One major legislative proposal seeks to promote sharing of information about computer security risks (often a good thing in assuring security over time) by creating a new exception to the Freedom of Information Act for computer security information. As currently proposed, however, the Davis-Moran bill likely does not provide enough assurances that suppliers of weak security will have incentives to improve their security over time.

In summary, there are few readers who have previously immersed themselves in computer security, economic theory, military strategy, and the range of legal and policy issues discussed in this paper. A surprising, and I hope intriguing, contribution of this paper is to show a unified way for thinking about issues in these disparate domains. Experts in each field can benefit by testing the assumptions in their own field against the often-conflicting assumptions of the other fields. I believe considerably more work needs to be done, by experts in each of these fields, to understand how to assemble a better collection of technology, laws, and institutions for having the desirable degree of openness and hiddenness in the newly important field of computer security.

I.  Security and Obscurity

To develop our understanding of hiddenness and deception in computer security, we turn first to some simple examples about the use of hiddenness in the physical world. These examples provide analogies for when deception will help in the cyber-world. We will begin to understand why modern computer experts often say "there is no security in obscurity," while military strategists through the ages have often thought that obscurity -- hiddenness and deception -- was crucial to defense.

      A.      <u>Security and Obscurity in the Physical World</u>

The goal of defense is to keep the attacker "out" -- out of a fort, some other physical structure, or a computer system. If the attacker gets "in" -- to the physical structure or computer system -- then the attacker can cause damage or even seize control. In the physical world, the defender also usually cares about offense -- about the ability to go outside of the walls and harm the attacking army or strike at the attacker's base. But in computer security, for purposes of our discussion, the focus is entirely on defending the fort or computer system. There may come a time when offense is a more important aspect of computer security, but for now system administrators are far more concerned about keeping the bad guys out. One reason for this focus on defense is that it is so hard to know who is really attacking. For instance, if I send a virus back to infect the attacker, it is quite possible I am sending the virus to an innocent system that has been hijacked by the attacker. If so, that innocent system may be wiped out without any harm to the real attacker. Wiping out innocent systems could lead to serious liability problems (if it is a commercial system) or diplomatic problems (if it is a government system). It is also impolite and inconsiderate, to say the least. So most system administrators stay focused on defense.

Some kinds of defense are not subtle at all. The simplest way to defend in the physical world is to build a wall so high and strong that the attacker cannot get over it or through it. This big wall is the opposite of deception. Attackers can see the wall from far away. They can gauge the strength of the defenses and decide whether it is even worth trying an attack. The attack, when it comes, is likely to rely on "brute force" -- a term worth remembering when we consider computer security. Brute force attacks on castles include, for instance, a ram that knocks down the gate, an artillery barrage that knocks down the wall, or a rush by overwhelming numbers of attackers to put up ladders and climb over the wall. Of course, deception does occasionally help even against an enormous wall, usually when someone from the "outside" attackers gets "inside" the wall by trickery. Having a spy open the front gate is very useful. So too, in the most famous siege in history, would inserting a Trojan horse, once again leading to the opening of the front gate. These examples of deception reinforce the wisdom of computer security experts who concentrate their efforts on hacking by employees and other insiders, in addition to the efforts to keep the outsiders out.

Here is another kind of defense that is not subtle. Imagine two armies fighting in broad daylight on a flat plain, perhaps in Central Europe or the steppes of Russia. The attacking army looks ahead and sees an enormous number of tanks in front of it. Once again, deception is of little use. The attacking army either has more tanks, enough to attack, or it does not. When the attack comes, both sides can see the advance. If the attacker has enough "brute force" at the point of attack, then it may win. Otherwise, the defense will hold. In the absence of a literal smoke screen that blocks the view, deception will matter little to the outcome.[2]

Other kinds of defense rely on deception. My usual example is the pit covered by leaves, with the sharpened stake at the bottom. If someone is trespassing near my house at night, then this sort of pit in a pathway may be very effective. The attacker is walking along and suddenly falls into the pit. Deception has succeeded.

This example of the pit can illustrate many of the points included in the formal model in the next Part of the paper. First, note that the pit works far better against a first-time attack than a repeated attack. The first time the trespasser comes down the path he or she might easily fall in. If the trespasser somehow avoids falling into the pit the first time (perhaps jumping safely to the side as the leaves begin to collapse), then that trespasser will not be dumb enough to "fall" for the same trick a second time. For instance, the trespasser might get past the pit, break into my house, and go back out on the same path. The trespasser will know to step around the dangerous spot.

Knowledge about the pit can spread even if the first attacker is unlucky enough to fall into it. For instance, the attacker might be part of a group, and the others quickly learn to walk around the pit. The attacker might be alone, but may survive and radio out to friends of the existence and location of the pit. These friends might tell the whole world where the pit is, perhaps by posting the information to a bulletin board on the Internet. In these ways, the first attack transforms into a repeated attack. The next attackers learn how to avoid the pit. Put simply, deception worked better against a first than a repeated attack.

Defenders who realize this can try to make each attack seem like a first-time attack. One trick would be to dig multiple pits. The second attacker might make it past the first pit but fall into the second. The second attacker would not know the location of the second pit, and so it would seem like a first-time attack. One thinks of a minefield, where the attacker must get past one hidden mine after another to get to the objective. This multiple-pit approach is more secure than the single pit, recalling the security maxim of "defense in depth."

The defender might do even better. Suppose now that the attacks come every few days. In between attacks, the defender might unpredictably fill in some of the old pits and dig some new pits. This unpredictable or random action by the defense would render useless the knowledge gained by the previous rounds of attackers. Even if the attackers had a perfect map from last time, they would not gain any advantage this time because the hidden traps had moved. It would be a first-time attack once again, with deception helping the defender.

In these examples, the deception succeeds to the extent that the attacker does not learn from previous attacks how to get around the pits. One assumption here is that attackers learn from experience. The information about the pit somehow gets back to the attacking army, such as by observing where an attacker falls into a pit. We can now introduce an important distinction between the pit and hidden surveillance cameras used by the defense. The pit, once used, is visible. If my comrade falls into the pit, then I see the fall, hear the crash, and learn where the defense is so I can avoid it in the future. By contrast, I can readily carry out an attack and never see a hidden camera or microphone. The surveillance device does not reveal itself to the attacker. As discussed later, I can sometimes infer that a surveillance device exists. One example is if the defender later publishes pictures of the attack. But the fact of the attack often does not teach the attacker about how and where surveillance operates. Even repeated attacks won't reveal much if the defender keeps surveillance information confidential. This distinction between the pit and the hidden camera shows the variation in the extent to which defenses become known to attackers over time, and hiddenness is more successful the less that is learned.

---

[2]One can imagine additional facts that would make deception more helpful. For instance, the armies may suffer from lack of fuel or other supplies. Perhaps many of the tanks on one side are actually out of fuel, but are deployed in ways that make them look operational. Nonetheless, the point remains clear. When the two sides can see each other clearly -- have high-quality information about each other and know that to be the case -- then deception matters less.

It is not only the attacker that learns from experience. The defender also adjusts. For instance, attackers may become wary after a time about walking down a path. They might begin to use long sticks to feel out the ground before them and discover the leaves over the pit. In this example, perhaps the defender learns to cover the pit with something thick enough to withstand the stick but thin enough to collapse when a person actually steps in the right spot. Then begins another round of adjustment and counter-adjustment. Attackers might roll a heavy object before them, so that any thin covering will collapse under the roller and not under a person. The defender might then key collapse of the pit to an infrared detector, so that the person but not the roller will trigger a fall into the pit. The details of this adjustment and counter-adjustment may seem fanciful, but the main point is simple enough -- the attacks prompt a more sophisticated defense over time. Put slightly differently, the more sophisticated defenses are more likely to survive an attack. They will be more secure, and the process of undergoing attack will teach the defenders how to make their position more secure. Foreshadowing a later discussion, this is a key point of the open source movement. The process of testing forces a system to become more sophisticated and ultimately more secure. A claim of open source proponents is that the advantages from this testing process, in the form of increased security, are likely to outweigh any disadvantages. Much of the analysis of this article ultimately explores the conditions under which this claim is true.

B.  The Weakness of Obscurity and Deception in Modern Computer Security.

For all of this talk about hidden pits and attackers falling into them, key aspects of modern computer security have little truck with deception. The slogan "there is no security through obscurity" seems to members of the open source and encryption communities to be a truism. My point in this part of the paper is to show why obscurity is of especially little use today for three of the main areas of computer security -- firewalls, packaged software, and encryption. Each of these areas is characterized by repeated, low-cost attacks. As explained in the previous section, deception is especially weak under conditions of repeated attacks. The attacker eventually learns about the hidden pitfalls and is stopped instead by fundamentally strong security, such as a high, thick wall.

1.  <u>Firewalls.</u>  The analogy between a computer firewall and a medieval city wall is

straightforward. A strong barrier exists that allows friends to enter but keeps foes out. Either sort of defense can be set to various levels of security. In times of peace, a city gate may allow anyone to enter, with guards on hand to handle anyone suspicious. At a higher level of alert, guards might check the credentials of each person before entering the city. During a siege, the gates might be closed completely, barring all entry. Additional security might exist within the city wall. For instance, the armory (containing weapons), the mint (containing treasure), and the castle keep (containing the ruler) all would have additional protections against entry.

A company's firewall is similar.[3] For non-essential systems pretty much any message will be allowed entry. For secure systems, a password or other credential is required. Under severe conditions, such as a distributed denial of service attack, all messages may be blocked from entering the company's system.[4] Additional security will exist for priority functions, such as the system security (the armory), the corporate treasury (the mint), and the root directory (the ruler's residence).

In the physical world, an attack on a walled city is a major event. Attackers will try to batter down the gates or climb over the walls in the face of determined defense. Real people will die. In the prosaic terms of modern scholarship, this is a "high-cost" attack. By contrast, an attack on a firewall is often very low-cost. A hacker, located pretty much anywhere in the world, can send a probe into the firewall, looking for an opening. The hacker can send dozens, hundreds, or thousands of probes against a single firewall. If one door is left open the hacker can enter. Depending on the strength of the defenses, the hacker may then be able to cause damage or take control over part or all of the system.

To give a sense of what is meant by "repeated attacks," the U.S. military reports that its computer systems are attacked over 10,000 times per month. Even if a hidden trap works the first time or the first ten times, it will likely not keep succeeding against an adversary who attacks repeatedly.

    2. <u>Packaged software.</u> A next major topic of modern computer security is how to protect standardized software against hackers. Popular products, such as Microsoft Outlook, may be on millions or tens of millions of desktops. This sort of standardized software provides many advantages to users, including the ability to sit down at a new computer or another user's computer and be fluent already in how to use a program.[5] Standardized software, however, also offers an almost irresistible target for hackers. Hack one copy of Microsoft Outlook and you might learn how to hack all of the copies.

Designers of standardized software might try to use deception to stop these hackers. For instance, the designer might have the program freeze up permanently if a user hacked into inappropriate portions of the software.  This kind of defense would be similar to falling into the pit covered with leaves -- the attacker who goes to the wrong place never comes out again.

---

[3][Give textbook definition of "firewall."]

[4][Cite sources on DDOS attacks: (1) newspaper story on the attacks in early 2000; (2) technical article on what a DDOS attack is.]

[5][Cite Lemley article on ease-of-use for computer users.]

Deception will not work well, however, for protecting standardized software.  Suppose, for instance, that there are a dozen paths to hacking the code to do something forbidden such as send a virus or make illegal copies.  Suppose the designer puts traps on eleven of the twelve, to freeze up the program permanently if a hacker trespasses into the wrong part of the code.  Suppose, further, that the designer leaves the twelfth path free so that the designer can get back in to rewrite the code.[6]

This sort of defense would work reasonably well against a one-time attack.  In the physical world, an attacker would face a grave risk (11 out of 12) of falling into the pit and being injured.  Similarly, in the computer world, if a hacker can get only one copy of the program, and needs that program to keep functioning, then it will be too risky to fool around with the program and likely have it freeze into uselessness.  By contrast, a hacker who can buy (or illegally copy) many copies of the program will not mind much if one or a few copies freeze up.  This hacker can systematically try one possible attack after another until something works.  At the same moment, hackers all over the world can also try their favorite attacks.  Furthermore, the hacker community can communicate effectively about what they find on hacker bulletin boards or on the Internet generally.

This sort of hacker attack on standardized software is enormously different from the physical attack on a walled city.  To attack the city, an opposing army has to travel through enemy terrain and then overcome the challenge of the high, thick walls, all at a high financial and military cost.  To attack the software is something anyone in the world can do safely on their home or office computer  It can be a form of recreation rather than a risk to life and limb.  The defenses for standardized software thus must withstand repeated attacks, from people all over the world, in contemplation that the attackers can share information about which hacks are successful.

The importance of this information sharing is illustrated by the way that video game players today learn how to "beat the game."[7]  "Beating the game" is a (presumably) innocent version of hacking a software system -- users ultimately reach their goal of gaining control over the software.  An old-fashioned (although perhaps satisfying) way to "beat the game" is to keep trying by yourself until you overcome all the obstacles.  As an alternative, video game players today can also enlist a global network of fellow aficionados.  Web sites appear almost instantly after release of the game.  The sites offer "secrets" (press the third brick on the left to get the magic sword), "walk throughs" (on Level 13 here are the seven things you have to do before you attack the dragon), and even "cheats" (if you enter this code, your player will become invulnerable to all attacks and as strong as Superman). Translated back into the language of computer security, the video gamers share information, instantly and globally, about how to hack the software. A game designer might hope to rely on deception, e.g., there is a barrier on Level 17 that is very hard to get around. But if even one person on the outside learns the secret way around the barrier (or if a company employee "leaks" the secret to a friend) then the rest of the world may quickly learn about the secret and thus overcome the deception. Otherwise put, a lot of players try to "beat the game," they tell their friends how to do it, and they post their successes to the Internet.

---

[6]This sort of access by the designer is sometimes called a "trap door."  In this metaphor, the designer wants to get under the floor into areas that are usually closed to traffic, such as access tunnels to Mission Control.  To keep the subterranean metaphors straight, remember that using a trap door is helpful because it lets you get somewhere you wish to go.  But falling into a pit is not -- you can't go anywhere and you might fall on that sharpened stake.

[7]This paragraph is based on interviews with my sons Nathan and Jesse Swire, as well as occasionally over-long descriptions at dinner of how they beat video games.

There is an additional security challenge for makers of standardized software who wish to keep their flaws hidden. When patches are needed for flaws, then the company needs to inform a very large number of users about how to apply the patch. Imagine, for instance, if Microsoft tried to tell only "authorized service representatives" about how to fix a flaw in Outlook. There could be at least thousands and potentially millions of people who would need to put the patch into place. How likely is it that the company could get the computer security information out to all its authorized representatives without also having the nature of the security flaw leak out to unauthorized hackers? If the unauthorized hackers will get the information soon in any event, then elaborate efforts to prevent such leaks may turn out to be expensive but accomplish little. In other words, a system relying on hiding flaws will not operate well when very large numbers of people need to learn the secret in order to fix the flaws on each computer.

3. <u>Encryption.</u> Encryption is a third major area of modern computer security, along with firewalls and standardized software. The word "encryption" comes from the Greek word for "hidden," so it might seem exceedingly odd to say that being hidden does not work when it comes to encryption. Yet, in the sense used in this article, that is precisely the claim. The question, for this article, is whether deception paired with fundamentally weak encryption will succeed, or whether instead only strong encryption will provide security. My point for now is that deceptive tricks, analogous to the pit covered by leaves, will not work for encryption. The basic reason is that the attackers can attack the text of the message over and over again until they find a way in. Only a high and thick wall can work, not tricky pits covered by leaves. This explanation of encryption and deception will set us up for the theoretical model in the next section about deception in computer security.

a. *A short primer on encryption.* For readers not familiar with encryption and the fervent public policy debates it has spawned, there are now excellent and accessible writings on the subject.[8] Here are some basic concepts for the less initiated. A sender named Alice wants to send a message to a recipient named Bob. A malicious person named Mallet may try to intercept and read the message. Alice and Bob don't want Mallet to read the message, so they change the "plaintext" of the message (the words that people can read) into a "ciphertext" (random-appearing characters that will seem like gibberish to Mallet). Bob, once he receives the ciphertext, changes it back into plaintext that he can read.

---

[8]For instance, Singh [cite] provides a very readable history of encryption in different periods in history, including the present. Steven Levy, <u>Crypto</u> (New York, 2001) focuses on the development of public key encryption and the U.S. policy and legal debates of the 1990s. A standard reference for people in the field is Bruce Schneier, <u>Applied Cryptography</u> (2d ed.) [cite].

A central puzzle in encryption is how Alice and Bob can agree to encrypt and decrypt the message (change plaintext into ciphertext and back again) while preventing Mallet from doing it, too. Experts have learned to provide Alice and Bob with a "cryptosystem" and a "key." The cryptosystem is a mathematical system that has a standard way to re-arrange symbols ("put every second letter in front of the letter before it") and substitute one symbol for another ("change each letter A into the number 1"). Historically, when encryption was used most prominently by diplomats and military forces, the users tried to disguise the cryptosystem – they didn't want outsiders to gain any clues from how the encryption took place. Today, by contrast, many Internet applications use the standard RSA cryptosystem developed by mathematicians Ronald Rivest, Adi Shamir, and Leonard Adleman. Having a standard cryptosystem has the outstanding advantage of permitting anyone to communicate with anyone else over the Internet using encryption. The security of the RSA cryptosystem depends on a mathematical algorithm that is easy to calculate in one direction (when one encrypts the message) but extremely difficult to calculate in the other direction (when an unauthorized person tries to decrypt the message).

The security of the RSA cryptosystem also depends on each user having a special key to turn ciphertext back into plaintext. The idea of a key is simple enough. Suppose that the cryptosystem turns each letter into a number, such as A=1, B=2, C=3, and so on. There are 26 possible starting points, such as A=25, B=26, C=1, and so on. In this example, the cryptosystem is a regular pattern for turning letters into numbers. The key is knowing how to begin the calculation, by knowing which number corresponds to the letter A. In actual cryptosystems, the key is a long chain of randomized numbers. Attackers who do not have the key then need to try every possible combination of numbers to see if the key fits the lock (decrypts this plaintext). Trying each of the combinations, which can easily number in the billions, trillions, and up, is called a "brute force attack." An attacker who can try every single possible key will eventually be able to read the code. The response by those who build cryptosystems is to try to make the number of combinations so large that no known computers can try all the combinations.

      b. *Encryption and deception.* Encryption becomes more important when you have to send lots of important messages through a channel where other people can see or hear the message. In times when the post was not secure, letter writers used encryption. In the days of the telegraph, many businesses used encryption to keep their commercial secrets away from the eyes of the telegraph operators. For radio communications, anyone with a receiver could hear the message. Most famously, German submarines in World War II used the Enigma system when radioing back to headquarters. Allied cryptographers learned to break the system after enormous effort, helping to win the war and more or less inventing the computer as a by-product.

The need for encryption is thus not new with the Internet. But the Internet has been accompanied by an enormous increase in the need for and use of encryption by ordinary people and businesses. The Internet is a famously "open" system. The message from Alice to Bob is typically routed through many computers on its way through the Internet. That evil person Mallet might be in control of any of those computers. Mallet might make a copy of all the messages coming through his system and then comb through the messages looking for any that have commercial, diplomatic, or military value. In response, Alice and Bob need to encrypt their important messages, containing credit card numbers, trade secrets, large transfers of currency, and anything else they don't want Mallet to read and copy.

How strong must Alice and Bob's protection be? That is a bit like asking how strong the lock should be on your front door. An inexpensive lock from the hardware store may stop the casual intruder but not the professional burglar. You might invest more in your security system depending on variables such as the risk of attack and the amount of valuables in the house. Turning to encryption, a weak system might be enough to keep a nosy sibling out of your diary or a casual intruder from reading your files at work. But as the importance of the information increases, users will want greater assurances that outsiders will not read their messages. For instance, companies may not trust the Internet for important commercial activities if they think that their competitors (or countries working to assist their competitors) will be able to read the messages. Ordinary users may not trust the Internet for e-commerce if they believe that attackers will steal their credit cards. The assurance of strong encryption, then, is widely seen as important to promoting the overall expansion of confidence in, and use of, an open system such as the Internet.

Encryption today is an odd combination of open and closed. Modern encryption uses the "open system" of the Internet, with "open" meaning that messages bounce through many different nodes of a network in a form where the intermediate nodes can read and copy the message. Modern encryption is also strikingly open in its standard cryptosystem. The RSA algorithm has been made entirely public and has been subject to vigorous attacks since its invention in the 1980s. In cryptography circles, this success against vigorous attack is a crucial indicator that a system should be considered secure.

On the other hand, the RSA cryptosystem relies on a well-hidden private key for each user. This single piece of hiddenness -- the private key -- is currently the principle and often the only real impediment to having Mallet read the message.[9] Reliance on this hidden key violates the usual technique of "defense in depth," but the advantage of this approach is that a person can send and receive encrypted messages with anyone in the world who also has a private key.

In summary, the hidden world of encryption turns out, upon inspection, to be extremely open when it comes to how modern encryption operates over the Internet. Attackers know what they need to do -- break the RSA algorithm -- but they don't know how to do it.

II.   A Model for What Should Be Open and Hidden in Computer Security

The discussion to this point has developed two somewhat contradictory intuitions. The first intuition is that hiddenness is often effective for defense in the physical world. An attacker may fall into the pit covered with leaves and an army might be lured into an ambush where it will be wiped out. The second intuition is that hiddenness appears to play much less of a role in the key tasks of modern computer security. For firewalls, packaged software, and encryption security experts appear to rely a great deal on open solutions to security problems. Outsiders who are knowledgeable about a field may often know as much or more about a company's firewall, packaged software, and encryption than the company employee who is deploying the system. Consider a small business on the Net using a standard software program or encryption algorithm. In such cases, the role of hiddenness or secrecy is at a minimum.

---

[9]Other impediments might exist, for instance, if the message is sent through a virtual private network or other mechanism that makes it hard for Mallet to read even the encrypted version of a message.

This section of the paper reconciles the two intuitions by presenting a model of where being open or hidden will be useful in computer security. The password stays secret, but the process is well known to many. Complete "openness" here means that outsiders and insiders know the same amount about how the computer system works.[10] "Hiddenness" exists where the defender knows something about the system that the attacker does not know. The model has a static and dynamic component. The static component concerns how useful hiddenness will be to protecting the system as it exists today. The dynamic component concerns how hiddenness will affect the performance of the system over time.

For the static part of the model, the main elements that determine the usefulness of hiddenness are:

(1) The extent to which the attacker is attacking for the first time vs. attacking repeatedly.

(2) The extent to which the attacker can learn from an attack.

(3) The extent to which the attacker can communicate this learning to other attackers.

As explained further below, these three factors can conceptually be collapsed down to the single factor of the extent to which there is a first-time attack.

For the dynamic part of the model, openness may lead to responses to attacks in ways that actually improve the security of the system ("security-enhancing effect"). More broadly, openness may make the system owner more accountable, not just for better security but for other purposes such as verifying compliance with applicable laws ("accountability effect").

Examples will illustrate how these factors can tip toward either openness or hiddenness in defending a computer system. An example that uses complete openness might be a firewall program, perhaps called "Openfirewall," that is sold off-the-shelf to many companies. Consider the elements of the model:

(1) Repeated attacks. Openfirewall will need to withstand repeated attacks from hackers. Any one firewall may come under repeated attack. Openfirewall will also be deployed on many different systems, once again predictably leading to many attacks.

---

[10]I assume, however, that system users keep passwords, keys, and similar authenticators confidential. "Openness" in such a case means that the system of security is known equally to outsiders, but the precise password or key is not.

-14-

(2) Attackers learn from an attack. On any one system, hackers can try one attack after another, waiting to see if any of them allow access to inside the firewall. If a door opens in the firewall, the hacker can tell.

(3) Attackers can communicate amongst themselves. Hackers who find a flaw in Openfirewall can gain "hacker glory" by telling fellow hackers how to conduct a successful attack. Hacker bulletin boards and other Internet postings can spread the word instantly and at close to zero cost.

(4) Security-enhancing and other accountability effects. The threat and reality of these hacker attacks will create incentives for the Openfirewall designers to de-bug their system. If a flaw is discovered by hackers, Openfirewall designers have an incentive to patch the flaw quickly to maintain the software's reputation. The net result of the hacker attacks may be to improve the quality of Openfirewall over time.

On the other side, hiddenness can be extremely useful where the elements point the other way:

(1) First-time attack. As shown by the pit covered by leaves, a hidden trap works especially well against a first-time attack. If the attacker for some reason can only attack once before the defense re-sets, then hiddenness may be a cost-effective defense.

(2) Attackers don't learn from an attack. If the hidden trap is deadly, then the attacker dies and has no chance to communicate back to his or her comrades. In essence, the next attack is still a first-time attack because the second attacker knows no more than the first. The hidden trap may work again.

(3) Attackers don't share information. Hiddenness will be more effective if information about a successful attack is not shared with others. For instance, an unusually talented hacker who finds her way past the hidden traps might be bribed by the defender not to tell how it was done. In such a case, it would still be a first-time attack for all the other hackers.

(4) No net benefit from system improvement. Suppose that the defenders have patched all of the holes they can patch, but there is still a flaw in the defense. Being open about the flaw will tell more hackers how to exploit the vulnerability, without suggesting any improvements in the defense. In the history of cryptography, for instance, there have been periods where the best decrypters could read anything sent by the best encrypters.[11] In these circumstances, being open does not result in system improvement,[12] increasing the likelihood that it is rational to keep aspects of system design hidden.

---

[11] Early in the Twentieth Century, for instance, until the invention of machine-based encryption such as the Enigma after World War I, the best codebreakers in Europe could fairly easily read the diplomatic dispatches of other countries. Singh. That situation could conceivably be repeated in the future, if mathematicians learned to factor very large numbers in a way that would make the RSA encryption algorithm weak. In such an event, the attackers would once again pull ahead of the defenders. It would then be more important to use hiding and deception in deploying encryption, because of the lack of any fundamentally strong algorithm.

### A. First-Time and Repeated Attacks

The example of the pit covered with leaves has shown the point that deception works better against a first-time attack than a repeated attack. A corollary of that point is that deception can work if repeated attacks are somehow transformed into a first-time attack.

This concept is well known to cryptographers in the form of the "one-time pad." Suppose that Alice and Bob communicate once a day. Suppose that each carries a device that generates an entirely random key each day, but does so in a way that Alice and Bob both know the key. One way to do this is if Alice and Bob each have a copy of a book that tells how to generate each day's key.[13] Each key is used to encipher exactly one communication. The next day Alice and Bob move onto a new key. Cryptographers consider the use of the one-time pad to be essentially unbreakable, so long as Alice and Bob properly guard their books. Each attack on the ciphertext is essentially a first-time attack. Progress in breaking the message for one day gives no advantage in breaking the message for the next day.

Using a one-time pad is impractical in large enterprises, such as an army or corporation, because of the difficulty of distributing the daily keys so that your friends can use the cipher but your enemies cannot. Nonetheless, the standard use of keys in modern public-key cryptography can be seen as a way to create a first-time attack. Suppose that Mallet discovers Bob's key, perhaps because Bob was careless and left a record of his private key laying around his desk at work.[14] Even if Mallet knows that Bob uses the standard RSA cryptosystem, and even if Mallet knows Bob's key, that still does not help Mallet read messages that Alice or other users receive. In public-key cryptography, it is a "first-time attack" when Mallet goes after each user's key. This is one way to see why keeping a key hidden is so powerful -- it transforms a repeated attack (Mallet attacking a cryptosystem employed by many users) into a first-time attack (Mallet having to break the key for each new user).

For defense of some computer systems, using hidden traps may work relatively well because they are truly first-time attacks. One example might be defense of a complex computer system, where an attacker would have to simultaneously take out several control centers in order to gain the goal of the attack. In this example, which might apply to the electric power grid in the United States or a major air defense system, the attackers may have no good way to tell ahead of time how the several control centers are inter-related. Only in the course of the actual first attack will the attacker learn whether it is sufficient, say, to knock out three out of five control centers, if done in a particular order. The attacker cannot simulate a full-blown attack in advance. Probes against part of the complex system will not reveal how the system will respond to a full-blown attack. In a case such as this, hidden defenses may function very effectively. Traps can be set for the first-time attacker, with substantial confidence that the attacker will not know how to avoid them. A goal of system design might thus be to make it difficult or impossible for the attacker to learn the nature of the defenses in advance of a large-scale attack.

---

[12]More precisely, the likely return to the system owner in terms of improved security is outweighed by the likely cost to the system owner of exposing a vulnerability. For instance, exposing the vulnerability may marginally increase the likelihood of a fundamental breakthrough in system security, but at a high cost in sustaining extra attacks in the meanwhile.

[13]To have strong security, Alice and Bob must keep the books highly secure. In World War II, similar books used by German submarines were occasionally captured, greatly aiding the Allied efforts to break the Enigma cipher. [cite Singh or Levy]

[14]Fortunately, sophisticated readers of this article would never their passwords or keys lying about in this sort of accessible way.

Designers of computer systems can also use strategies to transform a repeated attack into a first-time attack. Picking up a lesson from the one-time pad, defenders can regularly change their defenses. The pattern of pits covered with leaves can shift about. Passwords and cryptography keys can and should change regularly. Whatever defense exists at one period of time should quite possibly shift at another period of time. In economic terms, the defense should shift where the marginal benefit of the shift outweighs the marginal cost. The benefits here include the way that a shifting defense turns a repeated attack into a first-time attack. The costs are whatever burdens are needed to change the defense. For instance, someone must fill in the old pits and dig new ones. The computer systems designer must assure that the new system is at least as secure as the old one.[15]

A changed defense is likely to be especially effective if it employs randomness. Imagine, for instance, if the pits covered with leaves were scattered randomly across a terrain and somehow assumed a new, random pattern each morning. Nothing learned on the previous day about the distribution of pits would help the attacker on the next day. Randomness is a good design principle in choosing keys and passwords.[16] I suspect there are additional ways to employ randomness in computer security. For instance, consider a software package mounted on millions of desktops. Perhaps each copy should be randomized in various ways. That way, a virus or other attack on one copy of software will not necessarily succeed against another copy.

The use of randomness to reduce the effect of a virus or other attack is well known, to say the least, in protecting against biological attacks. It is known as the sexual form of reproduction.

---

[15]To illustrate the risks of changing a key or password, imagine that there is a currently functioning security system and Mallet does not know the key or password. Imagine, however, that Mallet intercepts the new key or password as it is sent to Bob. In this event, the costs of shifting to a new defense turn out to have exceeded the benefit of changing the defense.

[16]In the Enigma code, one mistake by the Germans was to depart from randomness. In setting the daily code, the German rule was not to allow one letter to substitute for a neighboring letter ("s" for "t" or vice versa) at an important stage of encryption. The reason was to prevent the opposing side from trying "easy" substitutions such as neighboring letters. In practice, the Allies discovered this rule and found it easier to break the daily code -- the German rule meant that the real letter "s" could not be represented by "r" or "t", thus substantially reducing the number of combinations the Allies needed to test. See Singh.

An important puzzle in modern biology is why animals bother with sexual reproduction when it would be so much easier to reproduce asexually. Sexual reproduction can take so much effort in finding a willing mate, and creates such a risk that an individual will never find a mate to reproduce. It may therefore seem more adaptive – more likely to succeed in reproduction -- to reproduce without need for a mate. In answer to this puzzle, biologists have recently stressed the role of randomness in resisting infection by viruses and other parasites. If a virus is perfectly suited to break through the defenses of one parent, for instance, the child may have resistance conferred in the random mixing of genes from two parents. A clone of the first parent would succumb, but a child of two parents may survive.[17]

Returning to computer viruses and similar attacks, I suggest that system designers explore a broader range of ways to use randomness and shifting defenses in system security. Perhaps standard software should more often be shipped in five or fifty or five million separate versions, each subtly different in various dimensions to reduce susceptibility to a standardized attack. Perhaps system defenses should automatically shift from one configuration to the next, perhaps at random intervals and in unpredictable ways. The advantage of this approach is that it turns repeated attacks into new attacks against the changed system. If a hacker stumbles into a flaw in the system, the same attack may not work against other copies of the software. A computer virus might not spread as fast.

One disadvantage is that there may be unpredictable downsides when software changes and then interacts with other software. Perhaps a standard software package is known to operate smoothly with a suite of other software. But the same software package, if it randomly changes over time, may have new bugs when operating with the other software. Another disadvantage of shifting the defense is the burden on authorized users. Ordinary users experience this when they are forced to change their current password (which they have *finally* memorized) to a new password every 90 days, as I was required to do in my federal government position. The optimal balance between change and stability will obviously change depending on context, but hidden aspects of computer defense will be much more effective the closer it is to a first-time attack.

    B.    <u>Learning from Attacks: The Distinction Between Surveillance and Other Defense.</u>

If attackers learn nothing from an attack, then hidden traps will work just as well for the second or nth attack as the first attack. Two lessons for defenders are to find ways to reduce the learning achieved by an attacker and to recognize the sorts of defenses for which there is little learning.

In terms of reducing learning achieved by an attack, a defender may be tempted to ruthlessness. In the physical world, an enemy scout who learns secrets may be killed or kept incommunicado indefinitely. In the computer world, it is worth considering what mechanisms

---

[17][Cite Glenn Reynolds law review article on parasitism and sexual reproduction as a reason for having elections and term limits; briefly discuss monoculture in agriculture and the advantages of a diverse seedstock]

would reduce the amount that an attacker learns about a system from attacking a firewall or reverse engineering packaged software. As one example, computer systems use "proxy servers" that prevent an outsider from learning the location of a user or a file that is inside a firewall.[18]

There is a continuum between the attacker directly perceiving the hidden defense and the attacker not learning anything about the hidden defense. In practice, however, we can simplify to two basic outcomes -- the attacker learns about the hidden defense or the attacker does not learn about the defense in the course of the attack and at most deduces the nature of the defense later.

The hidden pit illustrates a defense that an attacker can discover directly. If attackers poke the ground in front of them, they may collapse the leaves covering the pit without themselves falling in. In that event, they directly and authoritatively discover the hidden defense. They then can avoid the defense and communicate the nature of the defense to other attackers.

At the other extreme is an undetectable surveillance camera that watches the attacker. In this event, the attacker never sees the hidden camera. Attackers never learn about the hidden camera unless the defender somehow reveals that information. For example, the defender might post on the Internet a picture of the people carrying out the attack. In that case, the attackers would learn about the existence of the hidden camera. Based on the camera angle and background view, the attackers might also deduce the location of the camera. Future attackers might then have the ability to enter the grounds without coming into the camera's line of view.

Those involved in surveillance have long understood the importance of preventing the opposition from knowing the extent of their surveillance. For instance, Neal Stephenson organizes his masterful novel *Cryptonomicon* around precisely this theme. The novel retells the story of the Allies' decryption of the German Enigma crypto system during World War II. The strategic question for the Allies is how much to act on the secret messages they have decoded. For instance, if a convoy is crossing the Atlantic and the U-boats are poised to attack, should the convoy shift course? If the convoy does, then the Germans might deduce that the Enigma system has been broken, undermining the long-term ability to win the war. If it does not, then many people and boats will be lost. The novel describes elaborate efforts by the Allies to create cover stories for how they get useful intelligence, thus reducing the likelihood that the Germans will realize that Enigma has been broken.

This same theme of retaining a hidden defense was also crucial to the entry of the United States into World War I.[19] At a time when the Germany and the United States were officially neutral, the Germans sent the famous "Zimmerman telegram" to the government of Mexico. The telegram offered enticements for Mexico to ally with Germany against the United States, including promises of returning to Mexico territories that it held prior to the 1848 war. British intelligence decrypted the communication, but the intelligence agency was extremely loathe to reveal to anyone else that it had the capability of breaking German codes. British intelligence thus went through an elaborate, and successful, effort to make the leak appear to come from within the Mexican government. The Zimmerman telegram became public, speeding the entry of the United States into the war while retaining the British ability to read German communications.

---

[18][Give standard definition of "proxy server" and write a footnote of two or three sentences explaining how they work.]

[19]The account here follows Singh [cite].

Moving to questions of modern computer security, the point of this history is to show an important way that surveillance differs from the firewalls and packaged software that we have thus far considered. The analysis above has explained reasons why firewalls and packaged software are subjected to repeated attacks where the attackers learn how to avoid hidden traps. Attackers do not learn, however, about surveillance (unless they deduce its existence indirectly). The shift to a networked computer environment, characterized by the likelihood of repeated attacks, thus does not imply a similar shift in the optimal level of openness about surveillance as a security tool. In terms of assuring the security of a system, open source may be the appropriate approach for firewalls and packaged software but not for intrusion detection,[20] digital wiretaps, and other surveillance.[21] There may be other reasons to favor open source for surveillance, as discussed below, but the basic argument remains for keeping the "sources and methods" of surveillance confidential when shifting from physical to computer defense.[22]

C.     Communication Among Attackers and the Diffusion of Knowledge.

In the history of attacks in the physical world, it was often extremely difficult for one attacker to communicate back to allies about a weakness in the defense. The soldier who fell into a hidden pit would not have had a radio. The platoon that found a path through the defenses would have had no easy way to tell others how to do it. The army that came close to defeating a walled city might make a small breach in the wall, but the breach would be filled by the time reinforcements arrived. In short, the new attackers would essentially mount a first-time attack,

---

[20][Give standard definition of intrusion detection]

[21]The argument for openness in encryption is slightly different. The argument for openness in encryption is similar to the argument with respect to firewalls and packaged software, in that Internet users should be prepared for repeated attacks on ciphertext as a message is routed through the open Internet system. A difference arises because in some settings the best defense is to use a secret cryptosystem, where the attackers do not know the algorithm and other procedures for converting plaintext into ciphertext. For instance, progress on breaking the Enigma could only proceed after Allied spies learned enough details about the Enigma system that the mathematicians could hope to figure out the keys for each day. Singh [cite]. This sort of secret cryptosystem will not work for most Internet users, however. The benefits of having the Internet come in large measure from the large numbers of users who can engage in e-commerce and other communications. For these users to exchange messages securely, they must share a known cryptosystem such as the RSA algorithm. Modern encryption on the Internet depends on an open cryptosystem and hidden keys. Those who seek even higher security, and who are not confident enough that the RSA algorithm is unbreakable, might prefer a hidden cryptosystem and hidden keys.

[22][Cite to standard discussion of why sources and methods should stay confidential]

against which hidden defenses are especially effective.

In the computer world, by contrast, communication is instant and practically cost-free. If one hacker discovers a flaw in a popular software program, other hackers can mount the same attack against computers all over the world, even in the same day or hour. A virus can spread literally overnight. If a popular cryptosystem is suddenly broken, then the ciphertext of millions of people might suddenly become plaintext. This ease of communications among attackers is a crucial reason why the balance between hidden and open shifts with the shift from physical to network security. What used to be a first-time attack can now easily be a repeated attack by new hackers who have learned the flaws in the defense. Tricks and deception will work less well. Fundamentally strong security, which succeeds even against the attackers that knows how the defense works, is more important.

This "multiplier effect" of a successful hack helps explain some of the strong rhetoric about the threat facing network security.[23] Network security seems more brittle than physical security, because one successful hack can pierce the defenses of many installations rather than a single, physical wall. The multiplier effect also explains why it is so very difficult to assess the level of threat against computer networks. To illustrate, suppose that an enemy state has systematically learned how to hack many standard software programs in the United States. Then, a relatively small group of hackers might use these standardized attacks against thousands or millions of computer systems on the same day. If this scenario is plausible, then we really do need to avoid an "electronic Pearl Harbor" that waits just in our future.[24] By contrast, suppose that the multiplier effect is quite modest. Individual systems differ enough so that it takes skilled tinkering to break into each new system. In this event, the scenario of a massive simultaneous attack on all major systems is much less plausible. The effect of communications among attackers will vary considerably based on how easily the recipients of communication can mount their own attacks.

In assessing the change, a fundamental question is the rate of diffusion of knowledge about an attack.[25] From the point of view of computer security, the worst-case scenario is that there is instant and universal knowledge of a hack once one person does the hack. More optimistically, the knowledge of a hack in the head (or computer) of one person may mean that the secret gets out slowly or not at all. The presence of instant and cost-free communications means that knowledge of an attack *can* spread rapidly, but not that it necessarily will.

---

[23][cite Handel on multiplier effects and changing role of technology in creating such effects]["force multipliers"]

[24]Even before the events of September 11, 2001, the term "electronic Pearl Harbor" was often used to dramatize the threat to the U.S. critical infrastructure from computerized attacks. [cite]

[25][In economic terms, as discussed below, the question is the efficiency of the market for spreading information about an attack. Highly efficient market is instant (like a gas through a vacuum). Inefficient market means that a hidden aspect of defense continues to work.]

To assess the likelihood of rapid vs. slow diffusion of knowledge, a system designer should assess the threats facing that particular system. For some systems, one must plan for a determined, resourceful, and competent adversary. In time of war, for instance, one should assume that the enemy will go to great lengths to break your codes or pierce your firewalls.[26] In such instances, it is prudent to assume that any known flaw in your defense will also be known to your enemy. One must assume that there is "no security in obscurity" against that sort of determined foe.

On the other hand, many computer systems are not likely to face such a determined and competent foe. In such instances, the actual attacker could easily be ignorant of the state of the art. It would be instructive to see empirical research about the diffusion of knowledge about computer attacks and how common it is for particular attacks not to demonstrate knowledge of known flaws in a system. In the absence of systematic research, my interviews with network security experts suggest that the rate of diffusion is often surprisingly slow. For instance, many attacks are by so-called "script kiddies," who get a step-by-step script for how to mount a particular attack. If step 23 does not work, the "script kiddy" quite possibly will not understand what went wrong. The attack will fail, even though a more skilled hacker might easily have figured out a way to solve the problem at step 23.

The gap between skilled experts and "script kiddies" has implications for how defenders should spread information about new attacks. The point here is that there is practical, albeit limited, benefit of security through obscurity. Suppose there is a new attack against a popular firewall program. One implication of less-than-immediate diffusion is that it may be appropriate to delay notifying the world of the problem until there is a patch available. If diffusion were immediate, then one instance of a successful attack could mean that all systems using the firewall must be alerted immediately. The practice of security professionals, however, is that there is often time to develop a patch and then announce the problem.[27] A second implication of the varying skills among hackers is that there may be ways to communicate among security professionals without teaching the less skilled how to do an attack. Privacy and security expert Richard Smith, for instance, has made public a number of security flaws in popular computer systems. When he does so, however, he speaks at a high enough level of abstraction that the nature of the problem is communicated to professionals but the precise script for mounting an attack will not be evident to most hackers.[28]

In addition to the relative skills of hackers, the speed of diffusion will depend on the market, legal, and other incentives facing attackers. The creator of a new attack, depending on the context, may have reasons either to spread the knowledge or keep it secret. Hackers may wish to keep a secret when they benefit from knowing something that others do not. An attacker who gains money or other benefits from doing a hack may wish to keep a low profile and prevent defenders from creating a defense. On the other hand, innovators may wish to receive recognition for their clever new hack, as a resume item or to gain "hacker glory."[29] Innovators may also tell a small number of others about the attack, and then have the technique "leak" to a broader public.

---

[26]The now-famous example of these lengths is the creation of the large installation at Bletchley Park in Britain during World War II to break the Enigma code.

[27][Cite to the policy of CERT at Carnegie Mellon]

[28][Interview with Richard Smith]

Legal rules may have an important effect on the diffusion of knowledge about a computer attack. For instance, documenting that an attack is successful may constitute evidence of commission of a crime -- "here is how I hacked into the bank and stole $10,000." In such cases, the risk of prosecution discourages telling how the attack was done. A more controversial example is where the law directly forbids discussion of how a computer attack works. The "anti-circumvention" provisions of the Digital Millennium Copyright Act, in order to prevent copyright violations, place limits on the disclosure of some computer security information. This anti-circumvention rule can be understood as a technique for slowing the diffusion of knowledge about an attack. The anti-circumvention provision has been avidly defended and vigorously attacked on a number of different grounds. We will return to the anti-circumvention provision below, but for now I simply point out that legal rules can create incentives either to hide or disclose information about computer attacks.

[Perhaps insert here more discussion of the ethics of sharing detailed information about how to defeat computer security.]

In summary on the possibility of attackers sharing information about an attack, the possibility of cost-free communication in the Internet era increases the possibility of repeated attacks by those armed with knowledge from the first attack. More empirical work is needed on the actual rate of the diffusion of knowledge about how to conduct computer attacks. The actual rate of diffusion is likely to vary based on factors such as the nature of the threat to the computer system, the market incentives facing attackers either to hide or disclose their attack, and the legal rules that also create incentives to hide or disclose.

D. A Dynamic Model and the Accountability Effects of Openness

We have thus far looked at aspects of a static model for when being hidden helps with computer security. Having hidden code will help a particular security configuration to the extent that there is a first-time attack, or an attacker does not learn from previous attacks, or subsequent attackers do not learn from previous attackers. The next task is to look at a dynamic model for choosing between using hidden or open approaches to computer security. Advocates for the open source movement believe that having open code will improve security because the benefits of discovering bugs outweigh the security losses from openness. Even more generally, having open code may promote accountability of various sorts, much as having the Freedom of

---

[29]To give one example, the computer security expert Ian Griggs discovered high-profile security flaws in major software program while he was still a graduate student at Berkeley. Publicity about these skilled attacks helped develop his professional reputation. He shortly thereafter became Chief Scientist at ZeroKnowledge, a start-up company that attracted large amounts of venture capital. [cite]

Information Act promotes accountability of government. Over time, that is, openness may push systems toward having better security and meeting other goals that are associated with having openness.

1. *The security-enhancing effect.* Defenders of a physical or computer system face a difficult choice about whether to disclose the nature of the defense. Being open can lead to more effective de-bugging, but it can also be a roadmap for future attacks. Using secrecy can allow for hidden traps, but it also makes it more likely that serious flaws will be exploited when an attack actually comes.

The claim in this section is that the shift from physical to computer security makes it more likely, at least for the foreseeable future, that openness will be the optimal solution for enhancing security over time. Four factors contribute to this shift: the complex and buggy nature of software code; the likelihood of repeated attacks against computer systems; the importance of inter-operability to computer systems; and the likelihood that necessary expertise will reside over time with outsiders.

(a) *Many software bugs.* The presence of many bugs in software increases the likelihood that openness will be helpful. In theory, it may be possible to develop security for a computer system from a formal security model. This formal approach could, in theory, create secure systems with a minimal number of bugs. The experience of the past decades, however, is that this model-based approach to security does not work well.[30] Instead, modern computer security has a pragmatic, fix-problems-as-they-arise flavor. This pragmatic approach to security works has proven to be faster and less expensive than creating security from first principles. It is also more responsive to the practical challenges of creating security for complex systems, which must accommodate legacy software and hardware components.[31]

---

[30]Dorothy Denning has explained why pragmatic approaches to computer security have worked better than relying on formal computer security models. [cite] Denning stresses [say main factors in Denning]

[31]According to authors of a treatise on firewalls: "The major impediment to effective host security in modern computing environments is the complexity and diversity of those environments. Most modern environments include machines from multiple vendors, each with its own operating system, and each with its own set of security problems. Even if the site has machines from only one vendor, different releases of the same operating system often have significantly different security problems. Even if all these machines are from a single vendor and run a single release of the operating system, different configurations (different services enabled, and so on) can bring different subsystems into play (and into conflict) and lead to different sets of security problems. And even if the machines are all absolutely identical, the sheer number of them at some sites can make securing them all difficult. It takes a significant amount of upfront and ongoing work to effectively implement and maintain host security. Even with all that work done correctly, host security still often fails due to bugs in vendor software, or due to a lack of suitably secure software for some required functions." D. Brent Chapman & Elizabeth D. Zwicky, *Building Internet Firewalls* 14-15 (O'Reilly & Assoc., Inc. 1995).

The pragmatic approach to computer security also means that there is likely to be a significant number of bugs in a complex software package or computer system. The question for the system designer is then how open to be about the presence of bugs. The designer presumably will get "inside" testers to try to find and fix the bugs. If these inside testers are competent and given enough time to work on the system, then they will find a significant fraction of the bugs. Should the system designer then also open the system's code to outsiders, hoping to discover and fix additional flaws?

(b) *Repeated attacks on computers.* The answer depends in part on the likelihood that the system will be subjected to repeated attack. Suppose that you know for certain that there will be thousands of attacks against your firewall or packaged software. Attackers will presumably learn quite a bit about the hidden traps (that you inserted) and the hidden flaws (that you wish you knew about). Earlier attackers may communicate with later attackers, spreading knowledge about your vulnerabilities. In such a case, the designer may wish to create incentives for those who learn about attacks to report publicly about the flaws. The concern otherwise is that determined foes will learn about the attack but the system defender may not. For instance, if knowing how to hack the system is evidence of criminal entry into the system, then hackers will be reluctant to admit such knowledge, lessening the knowledge of the defender about flaws. If telling how to hack a system is illegal under "anti-circumvention" laws, then knowledge of how to hack the system will be forced underground. A better approach may be to publish openly the source code for the program, so that the discussion of flaws can also be in the open. By contrast, if one has few or no flaws in the defense, or only an occasional attack to face, then there is much less likelihood that outside review will actually improve the defense.

(c) *Security while allowing inter-operability.* The buggy nature of computer systems and the likelihood of repeated attacks thus tilt the optimal outcome toward openness. The tilt is increased by the importance of interoperability to modern system design. Take an example from the university setting. At my law school, grades are entered into the law school computer system. Grades are also shared for certain purposes with the main university administration. The interoperability challenge is how to share the grades securely while defending against the predictable attacks from student hackers who would love to transform their low grades into high ones. Suppose, for instance, that the university's software had been written in part by undergraduates from the engineering school. Is it possible that these undergraduates left a back door to allow them access to the grades? If so, should the law school trust grade reports from the university? Should the law school insist on seeing the code used by the university to determine whether it is trustworthy?

This example illustrates how others may insist on openness if my software or hardware is going to be used in their systems. They will need to know details about how my software or hardware operates. The more they rely on my software, the greater their security risk if they don't check my software before installing it.

It is possible that some products can be hermetically sealed in what the industry calls a "black box,"[32] which would be an independent sub-system that does not have negative effects on the rest of a system.  In such instances, the recipient of the black box cannot see its inner workings and presumably does not need to.  In many situations, however, the other system owner does not want to rely on a black box.  The black box may degrade the system operation, such as when messages are delayed when they are routed through the box.  The black box may have unsuspected other effects on the system, such as when its software is not compatible with the system's other software.  The black box may also be distrusted by the system owner, who might for instance believe that proprietary information will be captured by the black box and revealed to others.

      For these and similar reasons, another system owner may prefer to understand my software rather than rely on my black box.  Other system owners might deploy my software only if guaranteed at least a certain level of openness.  These owners may insist on openness even if I say that it will compromise security.  They may do so because the other advantages of openness, such as optimizing their system and ensuring compatibility, are greater than the security disadvantages that come from disclosing hidden information.  They may also judge the security equation differently.  I will be concerned that openness will expose my product to new attacks.  For the other owners, however, openness also means that they can guard against security risks that my black box creates, and their security experts may insist on knowing the details of the black box and other important components of their system.

      (d) *Security expertise outside of the organization.*  The likely location of expertise is the fourth factor that tends to support openness in the pursuit of better computer security.  Suppose that Desi relies on a particular programmer or a particular vendor to develop a proprietary software package or computer system.  One day, the programmer might quit or suddenly be unavailable to work.  One day the vendor may go out of business or stop supporting the software.  In either event, Desi may be in big trouble.  There may be poor documentation of how the system works.  There may be complicated aspects of the system that will be difficult for a new programmer or vendor to figure out.  As explained by Chapman & Zwicky: "Some people feel uncomfortable using software that's freely available on the Internet, particularly for security-critical applications.  We feel that the advantages outweigh the disadvantages.  You may not have the 'guarantees' offered by vendors, but you have the ability to inspect the source code and to share information with the large community that helps to maintain the software.  In practice, vendors come and go, but the community endures."[33]

      By contrast, it seems much less likely that being open will improve security for physical installations.  To be sure, building a sturdy wall around a town or military base is no small feat.  Flaws in the defense, however, may be simpler for the defender to spot and fix than bugs hidden in millions of lines of code.  Repeated attacks on a physical base are much rarer than attacks on firewalls and packaged software, especially when that software is deployed at thousands or millions of sites.  Having inter-operability occurs occasionally for physical defense, but much less often than in a pervasively networked computer environment.  For instance, an army unit

---

[32][Give an example of black box.  Use discussion of whether Carnivore was a black box.]

[33]Chapman & Zwicky, supra note __, at 23.

might move through a sector defended by an ally. In such instances, the army unit needs passwords, maps of safe passages, and other information so that it can "inter-operate" with the defenses in that sector. In such instances, however, the defending army is likely to have direct and confidential contact with the allied unit that is moving through. The passwords will be given to these temporary "insiders", and will assuredly not be posted for the entire world to see. The magnitude of inter-operability for computer security is far greater. Thousands of other systems may need to use Desi's software, and many of those systems may never directly communicate with Desi about security issues. Desi may thus post the information openly to the computer security world, even though the defending army is more secretive about its system. Finally, concerning the location of expertise, the military has long practice at maintaining its own specialized technology. The expertise needed to maintain a fortress has historically been maintained "in-house" by the military or in the established world of military contractors, rather than being standard technology where outsiders are likely to be better than insiders.

In considering the four factors that suggest that computer security improvements will come from greater openness, it is possible that countervailing factors will become more prominent over time. For one thing, the relative efficacy of employee hackers may improve with time. Openness improves security only to the extent that persons on the outside find bugs that are not found by employees of the system itself. (Openness also harms security to the extent that the openness tips off outsiders how to hack the system.) Chapman and Zwicky may say that "the community endures," but the analytic issue is the extent to which "the community" catches bugs that employees of the firm do not. This is an empirical question whose answer may shift over time. The incentives and behaviors of outside hackers may shift -- even if "the community" provides helpful and low-cost de-bugging services today, it is uncertain how well future programmers will share their knowledge, especially if there is no clear market reward for spotting the bugs.

A related point is that computer security may mature as a technology, so that the rate of change and the need for outsider de-bugging may decline. Today, security professionals are trying to cope with rapid change in seemingly every aspect of system operation.[34] In the future, however, computer security may follow the path of many other industries, where practices become more established over time and there is less need for ad hoc patching.[35] If so, then improved in-house practices may mean that the marginal benefit of de-bugging from the outside may decline. [net cost of insiders (pay $ but have oversight) and outsiders (don't pay $ and maybe they steal)]

In summary, it is relatively rare in physical defense to believe that openness will lead to improved security. The design of military bases is usually considered confidential information.

---

[34][footnote Schneier congressional testimony early july 2001 on mounting complexity of security]

[35]A rough analogy is the shift early in the 20th Century from one-of-a-kind automobiles to the systematization of the Ford assembly line.

Banks do not tell outsiders the design of their vaults and burglar alarm systems.[36]  More generally, defense issues are pervasively handled as classified information, where information is shared on a "need-to-know" basis and with the precise deployment of defenses often considered top-secret.  There are strong reasons for believing that the four factors discussed here will mean that, for the goal of enhancing computer security, openness will far more often be the optimal strategy than for traditional physical security.  Changes in the practice of computer security may erode this conclusion over time.  One particularly intriguing issue, however, will be how to mesh the closed system for physical defense with the logic of a more open approach for the computer security that will increasingly lie at the heart of military operations.

     2. *FOIA and other accountability effects.*  To recap, the static model shows reasons why openness is more likely to be the optimal solution for computer security than for physical security because of the likelihood of repeated attacks, the ability of attackers to learn from previous attacks, and the ability of subsequent attackers to learn from previous attackers.  The discussion of the security enhancing effect shows additional reasons that openness is likely to provide additional benefits over time, such as the complex and buggy nature of software code, the importance of inter-operability to computer systems, and the likelihood that the expertise to update systems will reside over time with outsiders.

     The security enhancing effect is part of a broader effect of openness -- accountability over time.  The system designer may choose hiddenness as the best defense for the moment, believing that enough attackers will fall into hidden pits to make digging the pits worthwhile.  Over time, however, the hidden pits will become less effective as knowledge spreads about their existence and location.  Taking the longer view, the designer of a computer system may choose openness as an initial strategy, hoping that the benefits of debugging the system and informing outsiders about the system will outweigh the loss of use of hidden defenses.  Looking only at the effects on his own system, however, the system owner or software developer may not make this choice. The software developer, for instance, may not fully value the advantages of openness (or the disadvantages of a black box or other hidden code) to the other system owners that use his software or inter-operate with his system.  In economic terms, this advantage of openness may be an "externality."  Desi's cost/benefit analysis of openness may not include the benefits of openness to other system owners.  The next part of this article offers an economic analysis of externalities and other market failures that may lead to an inefficient level of openness in computer security.

     Achieving the right level of openness in computer security has other benefits in addition to enhancing computer security and reducing market failures.  The Freedom of Information Act (FOIA) in the United States exists in order to promote greater openness, democracy, and accountability in the federal government.[37]  FOIA is also a more general symbol, outside of the federal government, for the idea that greater openness promotes important values.  The potential importance of FOIA can be imagined if there were a proposal that "all computer security"

---

     [36]This statement is based on not-for-attribution interviews with persons engaged in bank security.

     [37][Cite to basis and purpose of FOIA.]

-28-

information should be exempted from FOIA due to a belief that public discussion of computer security would tip off hackers about how to harm computer systems. To go even further, the proposal might say that "all computer security" information was classified and that it was illegal to tell other people about flaws in computer security systems.

In this (presumably unlikely) scenario, it would become more difficult for interested groups and the general public to learn what the federal government was doing in the area of computer security. If the government happened to be doing a bad job of protecting government computers and safeguarding the nation's critical infrastructure, there would be no ready avenue for public knowledge and debate. It would be difficult for the public to assess whether the federal budget for computer security was too high or not high enough. Similarly, it would become more difficult for the public to hold the government accountable for Internet surveillance and other activities of the federal government that related to the Internet and high technology.

This scenario may seem far-fetched: FOIA exemptions for computer security, classification of information in civilian hands about computer security, and criminalization of actions that reveal computer security flaws. As Part V shows, however, less-extreme measures along these lines either currently exist or have been seriously proposed as a way to reduce the harms from computer hacking.[38] My point here is to highlight how such proposals can reduce the accountability of those involved in computer security. There could be an impoverishment of the public debate about computer security issues. The experts on the "inside" would be subject to less criticism from interested persons on the "outside." At a time when the Internet and computer security are clearly topics of great public importance, a better-informed public debate may well lead to better policy outcomes.

My point is not at all to say that the benefits of accountability always trump the needs for hiddenness in security. Instead, I am highlighting how FOIA values promote accountability and better debate about what values our society should institute in important areas such as computer security. Reasonable people will differ in each case about how to weigh the benefits of accountability against the risks from disclosure. Detailed study and debate is appropriate about how to construct the rules and institutions for deciding what is disclosed and what kept hidden. Nonetheless, if "computer security" becomes a basis for hiding large amounts of information from public view, then accountability and public values may suffer.

III. A Law and Economics Model for Assessing Openness in Computer Security.

To this point, the focus has been on the incentives facing the person who is defending and designing a computer system. The question has been to what extent hiddenness will work in the static and dynamic models. This Part of the paper puts forth a law and economics approach to

---

[38]Part V examines the Davis-Moran legislation to create a new FOIA exemption for "critical infrastructure" information, the anti-circumvention provisions in the Digital Millennium Copyright Act that often make it illegal to publish information about successful computer attacks, and the possibility of extending the military's classified information system to the civilian computer security sector.

-29-

the question, and asks more broadly what is the societally optimal level of openness and hiddenness in computer security.

In considering this question, I name the person Defending and Designing the computer system as "Desi." Desi may be deciding how open to be in defending a firewall, or may be the software designer for packaged software or an encryption system. I name the person who is External to that system, and who represents Everyone Else, as "Ethel." Ethel is deciding, for instance, whether she trusts Desi's software enough to incorporate it into her system without knowing how Desi's security software works.[39]

This Part first explains why an economic model is an accurate analogy to the model of computer security developed above. In this analogy, a defense that relies on deception or other hidden information is considered a monopoly, while an entirely open system is considered a competitive market. The discussion then turns to Desi's incentives when he is a monopolist, and shows why he is likely to disclose less than an optimal amount of computer security information. Moving beyond monopoly problems, there are other potentially significant market failures, arising especially from information asymmetries between Desi and everyone else. The incentives to be less than optimally open about computer security are even greater, likely, for government agencies than for companies that sell their products to other computer users.

To the best of my knowledge, no previous writer has made this connection between economic analysis and openness in computer security. The connection should be useful to the study of computer security, because it allows the familiar methods of economic analysis to reveal patterns in the (for most lawyers and economists) less familiar issues of computer security. The connection should be useful also for the study of economics, because the instincts that some information should be hidden for computer security suggest situations within economic theory when a free flow of information may not be efficient.

A. <u>Why An Economic Analysis is Appropriate.</u>

Three factors make it particularly useful to apply economic analysis to the issue of when information should be open or hidden in computer security -- efficiency, rationality, and market structure (that is, monopoly and competitive markets). Efficiency here refers to maximizing social welfare, rather than focusing only on the incentives of Desi, the system owner. Desi's action may impose a variety of costs on other, and may otherwise lead to less-than-optimal decisions about openness in computer security. Examples discussed below include possible monopoly power on Desi's part, but also a variety of other market failures discussed below.

---

[39]If these names, taken from the comedy show "I Love Lucy," do not seem appropriate then I invite readers to suggest alternatives. My choice of names builds on the standard use in cryptography of Alice who sends the message and Bob who receives it. Those names are drawn from the movie, "Bob & Carol & Ted & Alice."
The name "Desi" has the advantages of invoking a system "designer" as well as a system "defender." The name "Ethel" is obviously linked to Desi through the famous television show, and has the advantage of being an "external" person to the system, in the role of "everyone else" who is not privy to the hidden information about Desi's system. Even better, when she is doing helpful testing from the outside, Ethel is an "ethical hacker." (Brave readers can try saying "Ethel the ethical hacker" five times fast.) Before thinking of Desi and Ethel, I had considered using other short names such as David and Edna. Choosing standard names makes it easier to discuss hypotheticals about how a system designer and defender will interact with other computer users.

"Rationality" is an assumption in economic models and is also generally a good fit both for Desi and for Everyone else. We expect Desi as a commercial system owner to rationally seek to maximize profits. We also expect a military system owner to generally act rationally, using means such as computer security software to advance goals such as winning a war. As for the outside hackers, some will seek money by taking items of value from Desi's system. Competitors may seek to learn how Desi operates his system, such as when they wish to copy some good attribute of the system or when they wish to sell a similar feature to others. Military hackers may wish to gain advantages in some present or future conflict. Some outsiders will seek hacker glory, perhaps in order to enhance their reputation and later get a better job, perhaps for recognition within the hacker community, or for other reasons. In any event, it is straightforward to expect that Desi will have information that he will rationally seek to protect and that outsiders will rationally seek to discover that information.

The third salient aspect of the economic model for computer security is to see how the concepts of monopoly and competition map onto the question of whether security information should be hidden or open. I argue that keeping information completely hidden is analogous to Desi having a monopoly on that information. Desi knows things that no one else knows, and may get advantages compared with others based on that knowledge. On the other hand, completely open source information is analogous to a competitive market. Many programmers can read the code, for instance, and compete to develop an innovation that Desi and others may use in their systems.

I suggest that using these standard economic concepts of efficiency, rationality, and market structure can illuminate important aspects of computer security. For those knowledgeable about law and economics, using the language of economics turns the hard and obscure problems of computer security into far more familiar patterns. Now those trained in economics can have strong intuitions about whether information in computer security should generally be hidden or open. For instance, economic theory has a powerful presumption in favor of competitive markets. The analogy between openness and a competitive market is a powerful argument, in economic theory, toward favoring openness in the design of computer security. To the extent the analogy is accepted as valid, as I believe it is, economic theory pushes for a presumption of openness. In order to justify hiddenness in computer security, there needs to be a coherent economic explanation of why that hiddenness is welfare enhancing for society.

This question -- what justification is there for hiddenness in computer security -- turns out also to have interesting implications for economic theory. Standard economic theory supposes that reducing information costs (increasing information flows) tends to produce a more efficient and competitive market. There is one well-recognized exception to this rule of supporting greater information flows. In the area of intellectual property, economists and lawyers have

-31-

explained how reducing information flows can sometimes create an incentive for the production of useful goods and services. Notably, legal protection for trade secrets may be efficient.[40] If there were no protection for trade secrets, then firms would not invest enough in desirable innovations. Firms would fear copying of the innovation by others, leaving insufficient profits to cover the costs of innovating. More broadly, legal protections for copyright and patents create incentives to innovate and limit the ability of others to use information created by the innovator without the innovator's license. Applied to computer security, keeping information hidden is analogous to a trade secret. It may actually be a trade secret where the legal standard is met.[41] Desi knows things that he does not share with others, because he will profit by keeping the information hidden.

The interesting implication for economic theory, however, is that we can now see a second general exception to the usual economic rule that greater information flows are more efficient. This second exception concerns the particular *harms* caused by releasing information. One easy example is if a malicious person reveals my password, PIN, or encryption key to the world. Under the usual economic approach, more information is better in the sense of being more efficient. But there is a distinct efficiency loss if my password, PIN, or key is revealed. Criminals might use this information to steal money from me or get information that I did not wish to let them know. Criminals might use the information to pretend to be me, a situation known as identity theft. They might enter a computer system using a password, and cause any sort of damage or other trouble. Most often perhaps, the systems I use and I will need to spend time and energy changing passwords and doing whatever else is necessary to return to the status quo ante.

This effect is by no means limited to passwords. The more general way to put the idea is that greater information flows are inefficient when the extra benefits to competition from the information are outweighed by harms caused by the release of information. Information improves markets by sending more accurate signals about supply and demand. Information can also, however, be used to harm Desi's system or the systems of others. Hiddenness sometimes helps security because the advantages of releasing the information (debugging and the rest) are outweighed by the disadvantages (such as more successful hacks).

The interplay between computer security and economic theory thus provides us with competing intuitions that point in opposite directions. Openness in security is analogous to a competitive market, where outsiders have the same information as insiders. By invoking the competitive market, there is a stronger argument from economic theory for openness in computer security. On the other hand, the sometime usefulness of hiddenness in computer security highlights previously unarticulated implications for economic theory. To the extent we agree that hiddenness is efficient in a particular context, it reinforces the second general exception, previously unexplained, to the usual idea that greater information produces greater efficiency.

---

[40][Cite to Posner on law and economics of crime; also Bone and other law reviews on the economics of trade secrets]

[41][Standard cite to elements of a trade secret]

In deciding between openness and closedness, the facts will shift over time. One theme of this paper is that the change from physical security to computer security brings with it a shift toward openness as the efficient outcome for a larger proportion of information related to security. In the future, however, as technology and market structure change, this conclusion might shift. And it is to market structure that we now turn our attention.

B. <u>Market Structure, Monopoly, and Computer Security.</u>

In discussing the security-enhancing effect in Part II, the unstated assumption was that Desi was participating in a competitive market. When Ethel or another system owner considered inter-operating with Desi's system, she would demand openness about Desi's system in order to avoid the risk that inter-operating with Desi's system would lead to security threats, incompatibility problems, or other disadvantages to Ethel's system. Desi would supply openness up to the point that the advantages to Desi (such as greater sales or greater inter-operation with their systems) outweighed the disadvantages to Desi (such as lost security from revealing the location of hidden traps). I have explained why inter-operability is more important and more common for computer than for physical security systems. It is thus more often in Desi's self-interest to be open for computer than for physical systems, in order to convince the Ethels of the world to permit inter-operability.

Desi's incentive shifts if he is a monopolist rather than a participant in a competitive market. In the competitive market, Desi keeps supplying openness about his computer system until a competitive equilibrium is reached. That equilibrium will be where the total advantages of additional openness (greater usefulness to Ethel as system user and Desi as system supplier) no longer exceed the total disadvantages of additional openness (Ethel has less need for openness and Desi is revealing security information that helps attacks more than it prevents them). Under standard economic theory, this competitive equilibrium is optimal from society's point of view -- there is no alternative level of openness that produces greater benefits for Desi, Ethel, and all other participants in the market.

Standard economic theory also holds that a monopoly market results in a loss in social welfare. From the buyer side, Ethel will still use monopoly software even if she does not know as much about its security attributes as she would wish or a competitive market would supply. Ethel will pay extra, compared to a competitive market, to make sure that her system can inter-operate with a dominant system that has market power. From the seller side, having market power means that the amount of openness will depend on what is best for the security of Desi's system rather than the security of both Desi's and Ethel's system. Suppose, for instance, that an extra unit of openness will reduce Desi's security by $100. (That is, the openness will produce a predicted extra $100 in costs to Desi for protecting his own security.) Suppose also that there are 100 buyers who each will have to spend an extra $10 if Desi keeps the information hidden. Ethel and the others will spend more money, for instance, to make sure their systems are compatible with Desi's and to reduce the risk that Desi's system will cause security problems for their own systems. In this example, Desi will see $100 in costs from extra openness, while the buyers will see benefits of $1000 (100 buyers at $10 each) from that extra openness. In a monopoly market, it may be rational for Desi to keep the information hidden, even though from society's point of view it would be more efficient to have the information be open.

This example focuses entirely on the incentives Desi faces to protect his own system security. Another incentive for Desi is to protect his monopoly position. Desi may have trade secrets and other information that serve as a barrier to entry into the market. Releasing detailed information on system security may provide valuable tips about other aspects of Desi's system, helping competitors to enter the market. If Desi reveals flaws in his security, then competitors may use that information to attack Desi and try to sell competing products. In an uncertain world where Desi does not know what will erode his market power, Desi has an incentive to err on the side of keeping information hidden, even where the total costs of hiddenness on Ethel and other buyers exceeds the benefits to Desi of that hiddenness.

A third and related point is that the monopolist has an incentive to prefer a proprietary to an open interconnection approach. Note that "open" has two distinct senses. This article in general focuses on "open" information as the opposite of "hidden." Information that is "open" in this sense is known by outsiders. The second sense is where "open" is the opposite of "proprietary." A proprietary system may be "open" in the informational sense, because people on the outside can know how the system works. This proprietary system, however, may be protected by patents, copyright, or other legal rules. These legal rules mean that outsiders are only allowed to use or inter-connect with the proprietary system under certain conditions, such as with a license from the patent holder.

The market leader may prefer a proprietary approach due to network effects. An example from the early history of telephony can give a sense of the argument. At one point, the Bell System had roughly half of the telephone market, and many smaller carriers had market shares of a few percent or less.[42] The smaller carriers pushed for an open interconnection approach, so that customers of any phone system could call customers of other phone systems. The Bell System objected. It defended its proprietary approach to providing phone service, and opposed inter-connecting with other systems. This position made it far more difficult for the small phone companies to attract customers, because users wanted to have access to the large number of Bell customers. Telephone service is a classic example of a "network externality," where the benefits to users rise with the number of other people connected to a network. By refusing to inter-connect, the Bell System apparently speeded its passage to a dominant position in telephone service.[43]

In the telephone example, the use of proprietary technology made it more difficult for smaller companies to compete with the Bell System. It is possible that this proprietary technology may be entirely "open" in the sense that the information about it is known to outsiders. The patent system is designed to publish the information that is essential to the patent, and copyrighted information can ordinarily be read by others but not copied in violation of the law. It is also possible, however, that a company in the position of the Bell System would keep some information hidden about how its system worked. Some information might qualify as a trade secret under the law. Other information, even if it did not fit the legal definition of trade secret, might be information that would benefit competitors if known, and so the Bell System would prefer to keep it hidden. In this example, network effects meant that the Bell System could gain market power by refusing to inter-connect, providing the lure of greater market share as a reason to keep information either proprietary (known to the public but with Bell System control) or hidden (unknown to the public and thus with Bell System control).

---

[42][Check and rewrite as necessary this history.]

[43]A recent reprise of this debate is evident in the market for instant messaging, where AOL has been the market leader with its proprietary system and other providers have pushed for inter-connection. See [cite]

Recent scholarship has emphasized why this sort of network externality – where users benefit more from the presence of more users -- is especially likely to exist in computer markets. Mark Lemley, for instance, has explained why many sorts of software have network effects. Users, for instance, prefer not to have to learn multiple software programs. They prefer to use systems that can easily share files with each other, and they quite often use other people's computers (when traveling or working together on a project, for instance) so that they benefit from having familiar software on the new computer. For hardware, there are clearly network benefits from having inter-operability. Inter-operability may result from a standards process that allows all comers to hook their hardware to a system. Inter-operability, however, also sometimes falls prey to competition. A famous example was the fight between the VHS and Beta standards in the early days of the video-cassette recorder. Many technical experts viewed the Beta as a superior system, but it did not inter-operate with VHS systems and the latter won out in the marketplace.[44] In this example, the network effect eventually drove the market toward a uniform standard, but only after many Beta users had invested in systems that quickly became obsolete.[45]

This analysis of monopoly and network effects has shown three conclusions. First, the security risks to a monopolist of revealing information may outweigh the security benefits to Ethel and everyone else. In such a case, a rational monopolist will reveal less security information than is socially optimal. Second, revealing security information may pose risks to the monopolist of loss of market power. The loss of market power may outweigh the gain to the monopolist from releasing information, once again resulting in less information than is optimal. Third, the presence of network effects (which are particularly likely in computer markets) creates incentives for monopolists to gain market power by allowing less inter-operability and revealing less security information than is optimal. The second and third points are similar in that they involve the desire by the company to have market power. The second point differs, however, in emphasizing how release of security information can erode existing market power of a monopolist. The third point stresses how network effects create an incentive to act in a proprietary fashion and hide information in order to use the network effect to build market power.

These three incentives to reveal sub-optimal amounts of security information lead to the more general conclusions that we have reason to be skeptical about whether the optimal amount of security information will be revealed where there is monopoly power. In practice, this skepticism will come into play when a monopolist proffers various explanations about why it would be inadvisable to reveal more information. Observers will then face the difficult task of choosing between two plausible hypotheses. One hypothesis is that the monopolist is acting efficiently and is accurately describing the risks arising from revealing more information. Another hypothesis is that the monopolist is acting strategically, and is revealing information to a point that it is optimal for the monopolist but still sub-optimal for society. The second hypothesis, of sub-optimal disclosure, will be more convincing to the extent that facts are developed to support any of the three scenarios just described.

---

[44][Cite to this history]

[45]There may be an effect that points in the opposite direction from the view that computer markets are more likely to be monopolistic due to the presence of network effects. Some scholars have explored reasons why the Internet increases market efficiency, largely because of the lower costs of matching buyers and sellers. See, e.g., Froomkin and others on efficiency in law and economics.

      C.   Information Asymmetries and Other Market Failures.

The discussion of monopoly power focuses our attention on the possibility that the incentives facing Desi may lead to an inefficient outcome, such as where Desi hides security information for his own benefit but the benefits to others of revealing that information would be greater. Monopoly power is one example of the broader category of market failures. For computer security, that means that monopoly power is one example of a broader set of circumstances where the incentives facing Desi or others may lead to an inefficient disclosure of security information. We turn now to discussion of other possible market failures. This discussion will be illustrative rather than exhaustive.

Let me anticipate some possible objections before turning to examples of market failure. In discussing market failures, I do not mean that identifying a market failure implies the desirability of some government program to correct the market failure. Many scholars in recent decades have made the powerful point that the extent of market failures must be considered together with the extent of "government failures," or the various problems that can arise from an effort to correct market failures.[46] I accept this point. Where I identify a possible market failure in the disclosure of security information, the next step in the analysis is to examine whether the cure is worse than the disease. Careful attention should be paid to the relative strengths and weaknesses, for instance, of technological measures, market approaches, self-regulatory approaches, and government regulation, and I have attempted to do this in my previous writings.[47] This comparative institutional analysis requires careful attention to the facts in a particular setting. Depending on the context, I support a mix of technology, market, self-regulatory, and regulatory approaches.[48]

---

[46]*See, e.g.,* Neal Komesar, [book on comparative institutional analysis]

[47]*See* Peter P. Swire, "Markets, Self-Regulation, and Government Enforcement in the Protection of Personal Information," [cite]; Peter P. Swire, "The Uses and Limits of Financial Cryptography: A Law Professor's Perspective," [cite]; Swire & Litan, *None of Your Business: World Data Flows, Electronic Commerce, and the European Privacy Directive* (Brookings, 1998) at chapter 1.

[48]For instance, I am currently writing an article on "Reflections on the White House Privacy Office," which will explain my views on which institutions were appropriate for

An additional objection is that alarm bells go off for some readers when I discuss the category of information disclosures that causes "harm." Have I forgotten the First Amendment and the values it upholds? For instance, there have been concerns (perhaps most often from the right) that a wide-open Internet will result in a flood of pornography, defamatory speech, and piracy of intellectual property.[49] The response (perhaps most often from the left) has been that the First Amendment sets strict limits on government regulation in this area, and, even if it does not, that the better policy approach is to allow wide-open speech.[50] In this highly-charged debate on pornography and other issues, any suggestion that "information can cause harm" may seem to lend support to the side that favors greater government regulation of speech. On this view, even doing the analysis of market failures may support the opponents of free speech.

In response, I reject the idea that one should shy away from accurate analysis for fear of aiding those with whom one disagrees. Indeed, the First Amendment is justified in large measure by the precept that more truth and better policies will result when individuals are encouraged to state their views frankly. In discussing the harms that can result from flows of information, I am focusing in large part on behavior that would violate the law in the non-computer environment. For instance, there are longstanding penalties for fraud even though fraud is generally perpetrated through speech. There are penalties for stealing and other crimes, and for conspiracy to commit those crimes, even when the criminal activity is achieved through speech. If the mugger says "Your money or your life" and the victim hands over his wallet, there is no First Amendment defense that the mugger was exercising a right to free speech. In examining the various market failures here, I stress that I am not trying to take sides in the pornography or similar First Amendment debates that are hotly contested in the law of cyberspace.[51] Instead, I am trying to analyze the structure of markets that involve information about computer security. In the Information Age, as information becomes more powerful as a tool for offense and defense of important systems, it becomes more important to look at inefficiencies in the markets that govern whether the right information is hidden or disclosed. Understanding those inefficiencies will suggest the best legal and policy responses. Then, these possible responses can be assessed in light of the First Amendment and First Amendment values.[52]

---

addressing privacy issues during my time as Chief Counselor for Privacy to the Clinton Administration.

[49]*See, e.g.*, [georgetown early study on porn; check godwin for someone who wants strict enforcement of defamation on the net; microsoft on piracy]

[50]*See, e.g.*, [Ross, pornography; godwin, defamation; samuelson or littman, piracy]

[51][Cite to Reno I, Reno II/III, CIPA challenges, and academic writings including Catherine Ross]

[52]As a legal matter, the First Amendment applies to governmental actions that abridge free speech. [cite text] The values underlying the First Amendment, however, can apply to private actors. For instance, the importance of wide-open public debate can be used as an argument why a privately-owned discussion group should not be censored by the owner. The

1. *Malicious harm by a software writer.* To start with an easy example, it is possible that Desi maliciously wishes to cause harm to Ethel. For instance, Desi may create his software system in a way that helps him to steal from Ethel. One way to do this is to put a "trojan horse" or similar trick in his software program.[53] When Ethel uses the software, the hidden software code orders some action that harms Ethel. For instance, the software could order a bank to send $1 million from Ethel's account to Desi's. This would have the same effect as Desi forging Ethel's signature on a check, and would be clearly criminal behavior. Or, the software could order all of Ethel's hard drives to be erased, causing potentially large damages. This would have the same effect as a bomb blowing up Ethel's equipment, and again would be clearly criminal behavior.

It is simple even for non-economists to understand that this is inefficient behavior -- that there are net benefits to society from prohibiting stealing and intentional damage.[54] Even if Desi benefits (as he would from pocketing the $1 million), the losses to Ethel are at least as great. And, if we did not prohibit this criminal behavior and such behavior became more common, then the Ethels of the world would have to inefficiently spend more money and effort in guarding against stealing and bomb attacks. Put another way, there are enormous economic and other benefits from having a functioning legal system that punishes criminals, rather than living in a state of nature characterized by the war of all against all.[55]

2. *Information asymmetries and disclosure about software.* The example of the trojan horse illustrates how Desi's incentives may lead to less than optimal disclosure of information about how Desi's software system works. Even though Desi would benefit from receiving the $1 million, society is better off prohibiting that behavior, and criminal penalties against such behavior are appropriate. In what sense, though, should this sort of stealing be considered a "market failure"? Economists would say that Desi may be able to get away with using the trojan horse where there is an "information asymmetry", that is, where Desi knows information that Ethel does not. Software is known to be very complicated. For a number of

---

First Amendment would generally not apply to the private owner, but its underlying values would.

[53] The name of course comes from The Iliad, where the Greeks hid soldiers inside a large wooden horse. When the Trojans brought the horse within their city walls, the Greeks snuck out at night, opened the city gates, and conquered the city. As computer code, a "trojan horse" refers to software designed by Desi that Ethel does not realize has a dangerous function. Ethel uses Desi's software, and the code that Ethel does not understand does its job and harms Ethel's system.

[54]*See* Richard Posner, Law and Economics [get cite for economic rationale for efficiency of criminal prohibitions on stealing and battery/trespass]

[55][Citations to Hobbes, political philosophy, and the economic benefits of a rule of law] [One can agree with Hobbes that it is better not to be in state of nature; subsequent writers have largely agreed with Hobbes on that, but have disagreed on the best form of government, with most not supporting the strong monarch that Hobbes favored.]

reasons, the person who programs the software generally knows far more about its intricacies than the purchaser. It would not be efficient for every user to have to know as much as the programmer. Specialized programmers are more trained in the area, and gain economies of scale from producing one product that can be copied and used by many buyers. Indeed, once a software program is written and de-bugged, the cost for an additional copy of the program approaches zero, This allows purchasers to buy the program with a low investment of money and reinforces the information advantage held by the programmer over the buyer.

Where the information asymmetry is large, as it often is for software, then there is a greater likelihood that a seller can make the product in ways that may harm the buyer without the buyer being able to see the problem in advance. The flaw in the product can be malicious, as in the case of the trojan horse. The flaw may also be negligent, such as where the programmer believes that customers will fail to see the flaws, and so does not take reasonable precautions to fix flaws in the program.

Where information asymmetries exist, there are standard ways to try to reduce the market failure.[56] [The rest of this section will analyze the standard law and economics responses to information asymmetries. Without deciding which approaches, if any, are appropriate, policies to consider include:

o   Professionalization for computer security professionals, as with doctors and lawyers. Professionalization may help build a code of ethics for its members. Self-policing institutions may develop for breaches, similar to a medical board or state bar. Credentialing programs may help assure purchasers of a minimal level of quality (or, more pessimistically, they might reduce the number of suppliers and raise costs to consumers without improving quality).

o   Mechanisms for equalizing the information. Laws may develop that create incentives or requirements for the more knowledgeable party to disclose information. A software developer, for instance, may be responsible for certain categories of harm unless there is a clear warning of the risk.

o   Foster the development of third-party monitors for quality. Accountants, for instance, play this role in certifying that public companies meet generally accepted accounting principles. Accountants or other third parties might play a similar role in certifying security products as meeting minimum standards.

o   Liability rules for willful, reckless, or negligent security practices. Given the intense debates about existing tort rules, this topic is likely to be extremely controversial politically. In certain cases, however, the wrongful actions of a security professional may be egregious enough to support liability for the harm caused to others. A related legal topic is what effect to give to contracts where the security professional disclaims liability. In the history of American products liability, these sorts of disclaimers were given great weight until legal doctrine shifted in the 1950s and 1960s to permit plaintiffs to recover often even in the face of such disclaimers.

---

[56][cite to standard law and economics literature]

o  Other consumer protection laws. In other areas where significant harm can occur to ordinary individuals, such as consumer lending, there are consumer protection laws to require disclosures and provide a range of remedies against wrongful behavior. Over time, consumer protection laws may develop in the area of computer security.

These sorts of legal responses are put forward as possible ways to address information asymmetries in the market for computer security. In each context, the initial task is to determine whether the alleged market failure is substantial. If it appears to be, the next task is to analyze the likely government failures that would exist in trying to remedy the market failure. For instance, wide-ranging tort liability would act as a deterrence to companies supplying bad security. It might also over-deter supply of security products if sellers had to assume risks that were outside of their control, such as sloppy implementation by the system owner that bought the software.

The implication of the analysis at this time is not to set forth a comprehensive response to every possible market failure involving information asymmetries. The implication, instead, is that familiar tools of law and economics will illuminate what legal and institutional responses may be appropriate when we spot a likelihood of a large market failure in the degree to which computer security is open or hidden.

### D.  Incentives for Hiddenness in Government Agencies

[This section explains reasons for believing that an inefficiently low level of openness is more likely to exist for government computer security than for private-sector computer security. The enhanced tendency of government agencies to keep information hidden should be kept in mind when assessing claims that such information should properly be kept hidden.

o  Government agencies have less of a market reason to disclose computer security information. In a competitive market, software suppliers have a major incentive to be open about their code -- being open encourages purchasers to trust that the software has been well-tested, will be compatible with the purchaser's system, and does not contain malicious code. In a monopoly market, as discussed above, the monopoly seller has less of an incentive to be open because demand is not very elastic based on openness of computer security. Nonetheless, the monopoly seller will reduce its sales and profits if the code is so hidden that purchasers don't trust letting the software into their system or don't know how to make the software compatible with their system. For military and other government computer systems, even the monopolist's incentives to be open are lacking. Getting other systems to use the government's computer system is usually not an important priority. (And, where it is, the government may be able to require that the government contractor or ally use the government's system.) There is thus little or no built-in incentive for a government computer system to be designed open in order to encourage outside users to inter-operate with the government system.

- o  A military or other government system may also rationally face a different calculus concerning the advantages and disadvantages of openness. For some government systems, the loss from a single attack may be enormous. Enemies, for instance, might learn how to construct nuclear weapons or how to penetrate an important military installation. By contrast, many bugs in a commercial setting will result in modest losses from a first attack. The expected loss from a first attack can shift the balance between openness (more likely to have an attack soon but enhance security over time) and hiddenness (less likely to reveal a flaw now but less likely to enhance security over time).

- o  Institutionally, this possibility of a large first-time loss can tilt institutions toward being risk-averse when it comes to openness of computer security. If some of the flaws in a computer system can lead to enormous loss, then the system owners may develop institutions to guard primarily against these first-time attacks. The risk of a "leak" may seem greater in the government sector than in the civilian sector. People operating in the system may perceive a greater risk from a leak, without any corresponding incentive to have a gain from appropriate open sharing of information.

- o  The overall military strategy of the United States has been based in part on the philosophy of retaining a technological edge over adversaries. Given the likelihood that other countries will learn technologies over time, there is a game-theory reason for the U.S. military to slow the diffusion of information about its high technology. Greater openness in computer security will narrow the gap between the U.S. military (the monopolist on system information) and other countries. This incentive is similar to the monopolist who resists sharing information because of a fear of seeing its monopoly power erode over time.

- o  Because the government has an existing classification system, an institutional mechanism is already in place for handling sensitive computer security information in a way that enforceable keeps it hidden. In a corporate setting, with no similar institutional mechanisms and a built-in reason to share information at least with one's business partners, it is harder to set up expectations about such information should be hidden over time.

- o  Significantly, there have been repeated studies about whether government agencies over- or under-classify information. These studies have consistently found that agencies substantially over-classify information for a wide range of national security information. For computer security, where there are newly compelling reasons to be more open about security information, there is thus empirical reason for skepticism about whether government agencies will be as open as good policy would suggest.

IV.     <u>Sun Tzu, Clausewitz, and Deception in Military Strategy</u>

To recap the argument, Parts I and II set forth the reasons why the optimal level of openness in computer security will often be greater than for traditional physical security. Part III used a model from economic theory to analyze how hiddenness in computer security is similar to having a monopoly while openness is similar to a competitive market. This Part draws on another theoretical discipline with important insights for the study of computer security -- the realm of military strategy.

It is not hard to see the relevance of military strategy to the question of how to defend a computer system from attack. The vocabulary of "defending from attack" has an unmistakably martial tone, and military theory should presumably offer insights on how to ward off such attacks. Furthermore, a large number of computer systems are themselves military systems. Whatever the general conclusions about what should be open and closed in computer security, there will need to be more specific decisions about how open or closed to be for *military* computer security. Perhaps the most important reason for studying military strategy, however, may turn out to be the surprising nature of some of the results, and especially the way that important strands of military theory support a greater degree of openness in computer security than one might have expected.

The writings of Michael Handel, of the U.S. Naval War College, highlight the variation between what he calls "the two great theorists of war" when it comes to the usefulness of hiddenness and deception.[57] On one side stands the writer Sun Tzu, whose book *The Art of War* was written in China over two thousand years ago.[58] Sun Tzu has famously said that "all warfare is based on deception." On the other side stands Carl von Clausewitz, whose classic work *On War* grew out of the author's experiences in the Napoleonic wars. Handel shows how Clausewitz, in his comprehensive study of military theory, gave little importance to deception and keeping one's actions hidden. Clausewitz largely thought that deception would not succeed and would divert one's resources from the key to success -- the concentration of superior force at the decisive point.[59]

A. The Usefulness of Hiddenness Varies with Terrain and Technology

In considering the different views of Sun Tzu and Clausewitz, I find it helpful to imagine that each is correct, but each is thinking about a different terrain. For Sun Tzu, who emphasizes deception, imagine warfare in a mountainous area. Fog and clouds cover the region, so that it is often difficult to see the enemy. Ambushes are easy to arrange. The weaker army can defeat the stronger, such as when the latter is strung out on a mountain path and cannot bring its superior force to bear. In this world of mountains and fog one might readily conclude that "all warfare is deception."

---

[57]Michael I. Handel, Masters of War: Classical Strategic Thought xiii (Frank Cass, London, 2d ed. 1996). Although I have studied a good deal of the writings of Sun Tzu and Clausewitz, I wish to disclaim an expert understanding of the two authors or of the general field of military strategy. The discussion here relies on Handel for setting up the contrast between Sun Tzu and Clausewitz, especially in chapter 12 of his book, entitled "Deception, Surprise and Intelligence." Handel mentions the importance of terrain to how effective deception will be, but the working-out of the examples is my own work, as is the application to computer security.

[58]Sun Tzu, The Art of War (tr. Thomas Cleary) (Shambala, Boston 1988).

[59]"It is dangerous, in fact, to use substantial forces over any length of time merely to create an illusion; there is always the risk that nothing will be gained and that the troops deployed will not be available when they are needed." Carl von Clausewitz, On War, 203 (ed. and trans. Sir Michael Howard & Peter Paret) (Princeton University, Princeton, 1984). *See* Handel, *supra* note __, at 127 ("It must be remembered that Clausewitz's principal method of winning battles was through the concentration of superior force a the decisive point.")

By contrast, imagine a battle on the flat plains of Central Europe or the steppes of Russia -- the terrain that Clausewitz experienced during the Napoleonic wars. On these flat plains, it is difficult or impossible to hide one's army. In the time of Napoleon, transportation was slow, with armies relying heavily on foot soldiers and horse-drawn artillery. This meant that it was difficult to move quickly with a strong force, to outflank the enemy or feint in one place and then hit in another. Small wonder in these conditions that Clausewitz thought that feints were worth little. By contrast, a winning strategy under these conditions would be to mobilize overwhelming force at a key place. If the attacker could smash the middle of the enemy's line, then the defenders had to retreat or else suffer a devastating combined attack from flank and front.[60]

In addition to terrain, differences in technology can clearly alter the effectiveness of hiddenness and openness. Many innovations have reduced the effectiveness of hiding. Binoculars and telescopes can see troop movements far away. Infrared goggles help spot people at night. Radar can spot airplanes long before they reach the target (and stealth technology later emerges as a counter-measure). These are all changes in the ability of one side to perceive the activities of the other side.

There have also been tremendous innovations in communication. Telegraph lines helped the Union front communicate with President Lincoln during the Civil War. Radios and telephones transformed communications in the 20th Century. Computers are continuing that transformation today. Better communications, teamed with better perception of the enemy's actions, make it harder for the enemy to hide from you or surprise you. On the other hand, better communications, teamed with stronger weapons and airplanes and other speedy methods of attacking, make it easier to attack with force before an enemy can respond. Changes in technology thus have effects that make hidden actions potentially both harder to achieve and more powerful when they are successful.[61]

---

[60]For fans of American football, this strategy is known as "three yards and a cloud of dust." Use overwhelming force (in the form of big lineman and strong running backs), don't worry about trickery or fake plays, and simply run over the defense. This sort of strategy arguably was very effective in the days of Woody Hayes at Ohio State and a long string of Nebraska teams, but changes in rules to favor the passing game have made the strategy less effective. Once again, changes in the constraints (terrain, technology, rules of the game) vary the extent to which hiddenness and trickery are successful as a strategy.

[61]Writing so soon after the tragic events of September 11, 2001, the mixed effects of technology on hidden operations are painfully evident. As shown in the Gulf War, improved powers of detection and communication by the U.S. armed forces have made it difficult or impossible for conventional armies to oppose the United States. On the other hand, the sheer

B. <u>Lessons for Computer Security</u>

The difference between plains and mountains has a strong analogy to the realm of computer security. As discussed in Part I, the computer security topics of firewalls, packaged software, and encryption all currently involve a large dose of openness. These technologies are characterized by repeated and low-cost attacks. The attackers typically learn from each attack, and it is easy to share information about successful attacks with other attackers. Under these circumstances, a hidden trap is unlikely to be successful. Even if the first attacker falls into the pit (loses the phone connection or whatever), future attacks will simply avoid that particular pitfall.

Under these conditions of openness, analogous to the flat plains of Central Europe, it becomes more important to have fundamentally strong security. The strength might come in the form of a strong and high wall, or in a thicker concentration of infantry, tanks, or computing power. Napoleon is supposed to have been fond of Voltaire's aphorism that "God is on the side of the larger battalions." In the computer setting, one might say instead that "God is on the side of the faster microprocessors."

This approach is known as a "brute force attack" in the realm of encryption. In a brute force attack, the hacker tries every combination of keys until the correct one is found and the plaintext can be read. The possibility of brute-force attacks is what generated a large policy debate in the late 1990's about how long encryption keys could be in software exported out of the United States. Law enforcement agencies wanted to limit the length of the key, presumably to make it possible for the government's largest computers to read encrypted messages. Supporters of strong encryption, such as civil liberties advocates and many technology companies, wanted to use longer key lengths precisely to avoid having the messages read by the government and anyone else who had large computers.

[Additional points for this discussion:

o    The Pentagon reports over 10,000 attacks on military computers a month in the United States. In light of these repeated attacks, deception and hiddenness often simply will not work. The focus must be on fundamentally strong security, which will succeed even when the attacker knows the structure of the defense.

---

power and speed of modern technology, including civilian airplanes, means that a dozen or two hidden attackers can cause massive damage.

In the terms used in this article, the attack of September 11 was quintessentially a "first-time attack," for which hiddenness is likely to be especially effective. One likes to hope that there has been a strong "security-enhancing effect" in response to the September 11 attack. One question, which quite possibly will be the subject of future Congressional hearings, is the extent to which airport security would have benefitted from a greater degree of openness and peer criticism before the terrible attack.

-44-

o   One interesting and perhaps surprising point is that hackers, who often consider themselves sly and tricky, are on the side of Clausewitz, openness, and the brute force attack when it comes to computer security. Intelligence agencies such as the NSA and the CIA, which after all are bureaucratic organizations, are on the side of Sun Tzu in believing that deception is crucial and that, by implication, hidden traps are likely to be important to computer security.

o   There are other reasons within military strategy to choose between openness and hiddenness. For instance, under deterrence theory one often reveals information about one's defenses in the belief that this will deter attacks.

o   The conflicting intuitions about openness are especially important to the topic of openness in military computer security. The intuitions in favor of openness that come from economic theory and the analysis of computer security clash with the intuitions in favor of hiddenness that typifies military strategy. Yet the stakes are especially high -- what will produce better computer security over time. For procurement policy reasons and because of the improved quality of civilian manufacture, military computers also increasingly rely on off-the-shelf hardware and software. To the extent that the civilian sector relies on openness in computer security, military computers using the same components will need to develop strategies that can succeed in the face of this openness about the nature of their products' design and vulnerabilities. It is interesting to speculate whether this challenge facing the military may help explain the recent U.S. military interest in purchasing open source systems.

V.   Applications of the Theory

This Part applies the theory developed above to applications of computer technology (the open source movement), law (the Digital Millennium Copyright Act and its anti-circumvention provision), and institutions (whether there should be no classification institutions for civilian computer security). The actual degree of openness or hiddenness in computer security will depend on the combination of how technology, law, and institutions develop in this area. This Part also applies the theory developed in the paper to the Carnivore e-mail system and to proposed amendments to FOIA that would keep hidden some types of computer security information.

A.   The Open Source Movement and "No Security Through Obscurity"

[This section will describe the views of Linux supporters and the Open Source movement about the optimal way to achieve computer security. Within this movement, there is widespread support for the slogan that "there is no security through obscurity." The slogan is sometimes used to indicate that computers that are linked to the Internet can't hide -- one cannot achieve security by lurking and trying not to draw attention to oneself.[62] Attackers' tools can and will find the computer. Empirical support for this view comes from the experience of users who hook up for always-on home broadband service. My interviews with users suggest that there are often multiple attacks on a new user within the first day, showing that keeping a low profile will not be an effective security strategy.

---

[62][cite to firewalls text on this]

More importantly, the slogan is used to indicate the importance of opening up the code for a security system. The slogan has perhaps its strongest support in the encryption area, where standard texts now state that algorithms or other components of a cryptosystem should only be considered secure after they have been extensively tested in a peer review process. [add quotations from supporters of this view, such as Felt affidavit]

The battle between the open source and proprietary models has been dramatized in the business realm in debates between IBM, a Linux supporter, and Microsoft about how to build software for business uses, including security. One IBM manager, for instance, said that "The world has changed irrevocably with the Internet, and Microsoft's single-platform approach won't work. You have to be able to connect with things."[63] This imperative for inter-connection is one of the key reasons this paper has argued that we should expect a greater degree of openness for computer security than for physical security. In situations where a system manager has to rely on a complicated medley of hardware and software, the manager will wish to know a great deal about the system components in order to conclude that the overall system is secure.

Open source supporters also tend to place great emphasis on the security-enhancing effect of openness in computer security. [add quotes and explain in more detail the ways that Linux supporters believe that the open source process leads to greater security over time]

That said, the analysis in this paper suggests there are important limits on the slogan of "no security through obscurity." Notably, obscurity can be helpful indeed against a first-time attack, in situations where attackers do not learn from previous attacks, and in situations where other attackers do not learn from the success of a prior attack. On a practical level, many system managers add some non-standard layers of security to their open-source security system.[64] These tailored applications can make it more like a first-time attack when any particular hacker tests the system. Tailoring a defense also makes it less likely that news will spread about that particular systems's weaknesses -- script kiddies who can defeat a standard defense will likely not have received a script for the tailored defense. The tailored defenses may not prove invincible or even very powerful against a state-of-the-art attacker, but they may effectively foil the average attack.

"No security through obscurity" is a slogan, used by people who wish to emphasize the surprising degree to which encryption and other aspects of computer security must rely on openness. But a slogan should not substitute for careful analysis. In all of the ways suggested by this paper, obscurity can sometimes be helpful in the defense of a particular computer system. Sometimes obscurity helps, sometimes it doesn't, and it is the job of this paper to help understand when it does or does not.

---

[63] Adam Galenist, marketing manager for IBM Software for Linux in the European region. [cite]

[64] [cite]

B. <u>Anti-Circumvention and Legal Limits on Openness in Computer Security</u>

[This section will analyze Section 1201 of the Digital Millennium Copyright Act as an example of a legal limit on the disclosure of computer security information. Some key points:

o [Write introductory paragraphs on what 1201 does. It is an example of a legal requirement to hide computer security information and prohibit its discussion by the outside community. Some aspects of the section restrict reverse engineering and learning about the security flaws. Others prohibit spreading the word about how a flaw can be exploited.

o There is a computer security research exception to section 1201. It is very narrow, generally requiring permission from the vendor, which may be difficult or impossible to get if the vendor does not want to suffer embarrassment.

o Section 1201 only applies to "effective" computer security provisions. Courts, however, have found that remarkably weak security measures count as "effective" and that it is thus unlawful to discuss publicly the flaws in even these weak security measures.

o The Edward Felt case dramatizes the security problems that section 1201 can cause. [Describe the case.] Professor Felt himself, in his declaration to the court, effectively shows a number of ways in which section 1201 can chill the entire field of computer security research in the United States. [Explain.]

o One of Professor Felt's key points is the role of serendipity in advanced research. For instance, his research on sound waves can overlap with the research of seismologists on sound waves, and Professor Felt explains how current law may make publication of that seismologist's research a violation of section 1201.

o The potentially negative effects of section 1201 go beyond a general chilling effect on the field of computer security research. For instance, Professor Felt's research showed the weaknesses in digital water marking by a company called Variance  Suppose that the Pentagon wants to add water marking to some of its sound files transmitted over the Internet. That way, for instance, the Pentagon would be able to tell if a sound file had come from a Pentagon source. This kind of water marking might be useful, for instance, for tracking leaks and hacks. Now, suppose that the procurement office is trying to determine whether the water marking technology is effective. The company, Variance, may state that no one can erase the watermark -- after all, the recording industry has adopted it as the industry standard! Professor Felt may know that the technology is weak, but he is prohibited under the DMCA from telling the world about the weakness. (Even more plausibly, as Professor Felt explains, the research would never have been done because researchers would have realized that they would not be able to publish the results.) In this example, the DMCA has resulted in *weak* security being adopted by the Pentagon -- there are known flaws in the Pentagon's system, but it is against the law for researchers to tell the Pentagon about them.

o  The effects of predictably weakening security in this way, by suppressing computer security research, are less tolerable after the events of September 2001. As part of the re-examination of other laws in the wake of these events, it seems appropriate to revisit the anti-circumvention provisions in section 1201 to determine whether it unacceptably dampens computer security research and publicity about known flaws in computer security systems. For instance, Congressional hearings may be appropriate to determine how the limits on computer security research are operating in practice and the ways that the bans on public discussion of security flaws are affecting development of computer security.

C.  Institutional Limits on Disclosing Civilian Computer Security Information

As a next application, let us turn our attention to a proposal that I believe will seem logical to people experienced in military defense. The proposal would be to create a classification system for computer security, similar to the classified/secret/top secret system that exists for traditional military and physical security in the United States. Under this approach, information that would facilitate a computer attack would be considered "classified." Only those who have passed a background check would be permitted to see this classified information. Special sanctions would exist for anyone who leaked this classified information, ranging from loss of the security clearance to special criminal penalties.

This approach, of course, has been used previously even for areas of important scientific research. The most famous example is for physicists who worked on nuclear weapons projects. A large fraction of the cutting-edge research was considered classified information. Physicists who participated in the project made a bargain. They got access to the state-of-the-art data and theory that the other top secret scientists had discovered. In return, they promised not to publish and get public recognition for their own research into the top-secret topics. A similar bargain was offered to outstanding young mathematicians in the period after World War II. They could join the National Security Agency or similar agencies and get access to cutting-edge mathematics and cryptography research. But the price for working on the most interesting projects was that they had to keep their own research results secret. For both the nuclear physicists and the cryptographers, the attraction of working on the classified projects was enhanced by the possibility of acting patriotically to protect the United States against dangerous foreign foes.

A question now is the extent to which this model should be used as a way to create security for computer networks generally and especially for our critical infrastructure. In my view, there will undoubtedly be areas where the military or other government agencies will do classified research into computer security topics. To give one example, we would expect highly classified treatment of any systems that can authorize the launch of nuclear missiles or that allow the President to give military orders. The standard doctrine of "defense in depth" strongly counsels that these critical activities have layers of secrecy and protection around them, so that the successful hack of one part of a system does not allow an enemy to take control of the entire system.

That said, I do not believe the approach used for nuclear weapons and Cold War cryptography can be the general approach for computer security. One difference is that computer security involves a preponderance of civilian systems. In contrast to the World War II information about atomic bombs or the Cold War information about encryption, a large fraction of the expertise and personnel involved in computer security are outside of the military classification program. A related difference is that computer systems inter-operate. Few people outside of the nuclear weapons program needed to share software or data with the nuclear physicists. Few people outside of the National Security Agency received information about breakthroughs in encryption. By contrast, the security of data in one civilian computer system often depends on the security of that data when handled by upstream or downstream users. The security of one system often depends crucially on the security and effectiveness of software that comes from a business partner or an outside vendor.

The cost and clumsiness would likely be very great from creating a classification system for computer security professionals and information related to computer security. In the modern context, where every small e-commerce site and university department has to guard system security, the number of people responsible for security is large and growing. The federal process for doing background checks has already been swamped by the large number of clearances needed annually, and subject to criticism for being too slow and not doing a good job of screening risk.[65] If that system is greatly expanded to include a large portion of civilian computer security personnel, one might imagine long delays before clearances were granted. This sort of delay might easily reduce system security, if the costs of not letting personnel work outweighed the benefits that came from screening out untrustworthy personnel. Another weakness of a military classification system for computer security would be the cutting-off of information about system weaknesses. The analysis of this paper suggests, generally in accord with claims from the Open Source movement, that computer security will often be reduced if information about weaknesses is covered up. An additional challenge to any proposed classification system would be how to share computer security information outside of the United States, for an Internet where the same software and the same system vulnerabilities exist globally.

If the analysis here is correct, then the military classification model for civilian computer security will not be desirable. The implication is that a large portion of computer security activity will be staffed with ordinary civilian workers and information about system vulnerabilities will often be available in the unclassified realm. If this implication is correct, then we perceive another way of understanding why we should expect greater openness for computer security than we have had for physical security. The sorts of personnel controls that were central to the military classification system are unlikely to work well for the Internet, and different security models, based on the assumption of more open information about system vulnerabilities, will need to be developed.

At the margin, there may be useful measures to slow the leakage of information from trusted security professionals out to potentially malicious hackers. It may be useful to develop certification programs for security professionals, so that we have more confidence in reports of systems flaws that come from the certified professionals. It may make sense to develop more ways to report bugs without tipping off the script kiddies about a vulnerability. These sorts of measures should be part of the developing ethics and practice of how to share information responsibly without needlessly triggering attacks before patches are put into place. But these measures are likely to develop against a backdrop of greater information about system vulnerabilities than many system owners will wish had to exist.

---

[65][Cite to GAO and other criticisms of classified clearance process.]

>    D.   Technology, Law, and Institutions as Substitutes in Creating Openness and Hiddenness

[This section will expand on the ways the technology, law, and institutions interact to create an overall system that results in a particular degree of openness and hiddenness in computer security. The section will highlight conditions in which each approach is likely to be most effective at producing a desirable outcome.]

>    E.   Carnivore

[This section will examine the FBI's Carnivore e-mail surveillance system in light of the analysis of this paper. Key points include:

o   Provide background on how Carnivore (now renamed DCS1000) operates, as part of a broader suite of software for e-mail and web surveillance. On the reassuring side for civil liberties, it is important that Carnivore is used in the United States only pursuant to a court order signed by a federal magistrate. On the less reassuring side, there have been questions about the scope of the information gathered by Carnivore.

o   During the Carnivore hearings in the fall of 2000, there were calls from open source supporters to have open source for the Carnivore software. Reasons for supporting this disclosure included invocation of the slogan "there is no security through obscurity." The idea, apparently, was that Carnivore would operate more effectively once the outside community had a chance to critique the software and suggest improvements.

o   Based on this paper's analysis, it is doubtful that disclosing the source code for Carnivore would enhance security, that is, would enhance its effectiveness as a surveillance tool. The reason goes back to the model of computer security in Part II, where surveillance was distinguished from computer defenses. The distinction was based on the fact that attackers do not generally learn how surveillance operates, even where there are repeated attacks. So far as surveillance is concerned, a later attack strongly resembles a first-time attack. The rationale for the security-enhancing effect of openness, thus, does not apply. In addition, the FBI has stated that releasing the source code would allow individuals to evade the system. Even though this paper has also suggested reasons for looking at such claims skeptically (in Part III.D.), the claim seems plausible in this instance because of the

o   That said, there may be valid arguments for opening up the source code for Carnivore or other surveillance systems. The reason would not be based on enhancing the effectiveness of the system. Instead, the rationale would be that the gains in accountability from disclosing how the system operated (such as to prevent over-reaching of the system) outweigh any security loss from disclosing how the system operated.

o   Greater worries about Carnivore may exist where accountability and public reporting mechanisms are the weakest. For domestic uses, a federal judge generally sees the proposed information-gathering in advance and the target of the investigation almost always learns about the tap once the investigation is complete. For uses under the Foreign Intelligence Surveillance Act, which constitute a majority of Carnivore uses to date, there is less public accountability. The targets of surveillance, for instance, never learn about the tap. For uses of Carnivore outside of the territory of the United States, such as any tap done offshore on information entering or leaving the United States, there generally is no requirement of a court order or any accounting after the fact to the subjects of the tap.

### F.    Openness in Critical Infrastructure and Amending FOIA

[This section will examine what degree of openness is appropriate for critical infrastructure protection (CIP), such as the computers that operate the payments system and the electricity grid. Government facilities and regulators are often integrally connected to these critical infrastructure systems. I offer a condensed version of the discussion here.

Improving information sharing about computer security has been a significant aspect of efforts to protect the critical infrastructure, with the formation of the Information Sharing and Analysis Centers (ISACs) that promote information sharing in the financial, electricity, and other sectors. A related task has been to analyze the extent to which current legal rules or institutions create obstacles to the optimal level of information sharing. Congressmen Davis and Moran, for instance, have proposed creating a new exception to the Freedom of Information Act that would mean that defined "critical infrastructure" information submitted to the federal government would not be subject to disclosure under FOIA. The logic of this proposal is to promote greater sharing of information about computer security.

The analysis in this paper, though, shows reasons for questioning whether this proposed legislation would actually improve computer security or otherwise be desirable.[66] First, notice that the legislation would create greater openness among a defined community (the companies that share the information with the government) but would reduce openness for the same information in other respects. As currently drafted, for instance, companies would have large discretion to label information as "critical infrastructure" information and thus exempted from FOIA, even if the same information would have been shared with government in the absence of

---

[66]My view at this point is that there may be a better-crafted proposal that would enhance government computer security without the risks in the current Davis-Moran proposal. My analysis here builds on points raised in 2000 by a variety of people inside the government and others including James Dempsey and Patrice McDermott during consideration of the proposed legislation. After extensive consultation, the Clinton Administration decided not to adopt a formal position on the Davis-Moran bill.

-51-

the FOIA exemption.  This company discretion may mean that the net result of the legislation would be to reduce the openness of information about computer security -- information already shared with the government would enter into the realm of hiddenness, and this reduction in openness might outweigh any increase in sharing with the government.

Second, the current legislation grants companies an exemption from liability for critical infrastructure information they share with the government.  Liability for weak security is one current incentive for companies to create good computer security, such as where recklessly bad security by a company causes harm to other systems.  This category of security is part of the "security-enhancing" effect discussed in this article from having open information about computer security.  Reducing the accountability means that companies will be able to reduce their accountability for good computer security, and act below the standard that usually triggers responsibility for wrongful behavior, simply by deciding to send the information to the government labeled as "critical infrastructure" information.

Third, the analysis in the economic analysis part of this article showed reasons for believing that government agencies face particularly strong incentives to be less open about computer security flaws than is societally optimal.  To the extent this is true, then we should be especially cautious about legislating in ways that reinforce the hiddenness of computer security information in the hands of government.  Information that came from the private sector, which the government may now judge would be better shared with private computer users, would be perpetually kept secret under the Davis-Moran legislation with no apparent discretion by government officials to share the information without company permission.  An additional risk is that information could be labeled as "critical infrastructure" information, not for reasons related to computer security, but specifically to avoid the public oversight that generally comes with the Freedom of Information Act.

Legislative amendments might do a better job of encouraging sharing of computer security information with federal agencies without risking these sorts of perverse effects on computer security.  James Dempsey, for instance, has proposed tightening the procedures that would be needed to qualify for the FOIA exemption.  Notably, information might qualify for the exemption only if the government received the information as part of a broader sharing within an ISAC.  In this way, government would receive the same information as owners of private-sector computer systems, and government would not be disadvantaged in the protection of its own computer systems by the existence of FOIA.  But the existence of the FOIA exemption could not be used by companies as a way to shield themselves from the responsibility for good computer security that they otherwise would have.

This discussion of the Davis-Moran bill illustrates the complicated set of technical issues, laws, and institutions that feed into society's decisions about what should be the degree of hiddenness and openness for computer security.  Under this paper's general analysis, the optimal level of openness is often higher for computer security than for previous security issues.  But the analysis here also shows a number of situations where openness in computer security is likely not to be the best course.  In assessing the Davis-Moran bill or other proposals, there is no substitute for a step-by-step analysis of the effects of a proposal on computer security and other values over time.

VI.     Conclusion

This paper proposes that the topic of what is hidden and open in computer security deserves far more attention than it has received to date. When it comes to computer security, we have drastically competing intuitions. On the one hand, economic theory tends to predict that greater openness in the flow of information will lead to greater efficiency. In the area of computers, we tend to believe the Internet should operate under open standards, where the transparency of the standards allows everyone to connect to everyone else. Also, we know that computer systems are themselves complex and need to inter-operate with each other in complex ways. System owners will only wish to incorporate software and hardware that they trust (that they know enough about), and they will need to know a good deal about the other systems in order to inter-act effectively and securely. On the other hand, we tend to believe that security flaws sound like top-secret military information. Disclosing the flaws feels a bit like helping the enemy know how to attack. Surely the correct answer is not always to disclose everything about the security of one's computer system.

The approach in this paper has been to analyze the conditions under which each of these intuitions is most persuasive. Part II of the paper worked within the framework of computer security to show situations where hiddenness tends to be more or less effective. Part III used economic analysis, which generally assumes the efficiency of greater flows of information. That Part highlighted situations where monopoly power and other market failures will most likely lead to an inefficient level of openness or hiddenness. Part IV used military strategy to highlight the way that the effectiveness of hiddenness varies with the terrain and the technology.

One message of this article is that computer security is not simply a topic that should be left to the technical experts who can debate the stability of Unix versus NT or the settings for intrusion detection systems. Nor should decisions be left only to the vendors and buyers of software and security systems, because there are likely to be significant market failures that will result in the wrong levels of openness and hiddenness. In addition, there are risks to leaving the decisions exclusively in the hands of security experts in the classified community. In my experience, these experts are often extraordinarily talented and committed individuals. But repeated studies have shown that there are powerful incentives that lead to over-classification of information. The analysis in this paper suggests that the incentives for government agencies to hide computer security information may be particularly potent.

The theme of open and closed has run through many of the important debates about the Internet: fair use versus strong copyright rights; open access rules for broadband systems versus free market models; wide dissemination of personal information versus privacy protection; open source code versus proprietary software. The Internet is, in part, a great engine for openness because it allows the sharing of information in so many ways to so many people at such low cost. Getting the correct level of openness for computer security is important, because otherwise the usefulness of the entire Internet can be greatly reduced. If openness goes too far, and hackers make the entire Internet the equivalent of a free-fire zone for muggers, then fewer people will use the Net. If hiddenness goes too far, and no system owner knows when to trust other system owners, then the open inter-connection that defines the Internet can be lost. Somehow, to keep the Internet's benefits to education, commerce, democracy, international understanding, and all the rest, we must find out how to build technology, laws, and institutions that create the right mix of openness and hiddenness in computer security.